\documentclass{aa}  

\usepackage{graphicx}
\usepackage{txfonts}
\usepackage{natbib}

\usepackage{float}

\usepackage[bookmarks=false,colorlinks=true,linkcolor=black,citecolor=cyan,filecolor=black,urlcolor=cyan]{hyperref}
\usepackage[symbol]{footmisc}
\usepackage[usenames,dvipsnames]{color}

\begin{document}

   \title{The global energetics of radio AGN kinetic feedback in the local universe}

   \author{Z. Igo
          \inst{1,2} \&
          A. Merloni\inst{1}
          }

   \authorrunning{Z. Igo et al.}
   
   \institute{Max-Planck-Institut für Extraterrestrische Physik (MPE), Giessenbachstrasse 1, 85748 Garching bei München, Germany\\
              \email{zigo@mpe.mpg.de}
         \and
             Exzellenzcluster ORIGINS, Boltzmannstr. 2, 85748, Garching, Germany\\}

   \date{Received XX; accepted YY}

 
  \abstract
   {AGN feedback is a crucial ingredient for understanding galaxy evolution. However, a complete quantitative time-dependent framework, including the dependence of such feedback on AGN, host galaxy, and host halo properties, is yet to be developed.
   }
   {Using the complete sample of 682 radio AGN from the LOFAR-eFEDS survey ($z<0.4$), we derive the average jet power of massive galaxies and its variation as a function of stellar mass ($M_*$), halo mass ($M_h$) and radio morphology.
   }
   {We compare the incidence distributions of compact and complex radio AGN as a function of specific black hole kinetic power, $\lambda_{\rm Jet}$, and synthesise, for the first time, the radio luminosity function (RLF) by $M_*$ and radio morphology. Our RLF and derived total radio AGN kinetic luminosity density, $\log \Omega_{\rm kin}/[\rm {W~Mpc^{-3}}]=32.15_{-0.34}^{+0.18}$, align with previous work.
   }
   {Kinetic feedback from radio AGN dominates over any plausible inventory of radiatively-driven feedback for galaxies with $\log M_*/M_\odot > 10.6$. More specifically, it is the compact radio AGN which dominate this global kinetic energy budget for all but the most massive galaxies ($10.6 < \log M_*/M_{\odot} < 11.5$). Subsequently, we compare the average injected jet energy ($\overline{E_{\rm Jet}}$) against the galaxy and halo binding energy ($U_{\rm bin}$), and against the total thermal energy of the host gas ($E_{\rm th}$) within halos. We find that compact radio AGN lack the energy to fully unbind galaxies, but complex AGN reach $\overline{E_{\rm Jet}} > U_{\rm bin}$ in the most massive systems ($\log M_*/M_\odot > 11.5$), where such energy is likely deposited beyond the typical galaxy sizes. On halo scales, neither compact nor complex radio AGN provide enough energy to fully disrupt the \textit{global} gas distribution, especially not for the most massive clusters. On the other hand, $\overline{E_{\rm Jet}}$ greatly surpasses the global $E_{\rm th}$ for groups, thereby providing a crucial input to the gas and thermodynamical balance in these systems. Finally, we show that AGN jets can also significantly impact the \textit{local} thermodynamical balance in the cores of large groups and massive clusters. Overall, our findings provide important insights on jet powering, accretion processes and black hole-galaxy coevolution via AGN feedback.}
   {}
 

   \maketitle

%

\section{Introduction}
\label{sec:intro}

The importance of active galactic nuclei (AGN) feedback in galaxies, groups and clusters has been empirically elucidated over the last two decades \citep[e.g.][]{Fabian2012, 
King2015, Eckert2021, Hlavacek-Larrondo2022, Bahar2024}, and this phenomenon is now a required ingredient in our theoretical understanding of galaxy formation, as demonstrated by several cosmological simulations \citep[e.g.][]{Croton2006, Bower2006, Hopkins2006, Sijacki2007, Schaye2015, Croton2016}. However, we still lack a quantitative understanding of the different modes of feedback, of the dependencies on AGN and environmental parameters over different scales and cosmic times, and how these should guide fine-tuned inputs to simulations \citep[e.g. see recent reviews by][]{Hardcastle&Croston2020, Harrison2024}.

Classifying in broad terms, AGN feedback can be either `radiative' or `kinetic', depending on the main channel of AGN energy release. The former pertains to growing black holes releasing energy (mainly) through photons, thus producing powerful radiatively-driven winds; the latter indicates those states or phases of AGN in which most of the energy is released as particles' (often relativistic) bulk motion, such as in jets \citep[e.g.][]{Begelman1984, MerloniHeinz2007, Alexander&Hickox2012,HeckmanandBest2014,  Harrison2018, Blandford2019, Costa2020}. Both feedback modes can have positive and/or negative impact on the black hole growth itself, on the structure of the host-galaxy or even on the larger scale host-halo. For example, suppression or enhancement of star formation, depletion or replenishing of gas reservoirs, halting or fuelling the growth of the supermassive black hole (SMBH) itself \citep[e.g.][]{Ferrarese&Merritt2000, Bower2012, Kormendy&Ho2013, Harrison2017} are all secular phenomena in the evolution of galaxies linked to AGN feedback phenomenology, and are known to act on varying timescales. These aspects make it challenging to accurately disentangle causality from effect, to quantify the energetics of a given mode of feedback and to understand on what scales each of these modes has the greatest impact, not to mention how this varies with cosmic time. 

Nevertheless, one aspect that is now observationally clear, and relevant in the context of this work, is that the kinetic power attributed to particle jets can efficiently offset cooling flows in galaxy groups and clusters \citep[e.g.][]{Fabian1994, Peterson2004,McNamara2005,McNamara2012}, thereby preventing extreme star formation in the massive central galaxies of those halos (note that gas does not necessarily have to be expelled from the host galaxy or halo to be considered `negative feedback'). In fact, bubbles inflated by jets have been found to be coincident with X-ray cavities in clusters, showing direct evidence for AGN kinetic feedback and allowing estimates of the jet power to be computed \citep[e.g.][]{Boehringer1993, Carilli1994, Birzan2004, Dunn2006, Cavagnolo2010, Timmerman2022}. However, the efficiency with which jets of varying power, orientation and morphology couple to the gas in the surrounding inter- or circumgalactic-medium (ISM, CGM) is unclear \citep[e.g.][]{Wagner2012, Mukherjee2018, Perucho2019, Meenakshi2022, BourneYang2023}.

One approach to gain empirical insight on AGN feedback is to use large, well-characterised samples to trace the global properties of a given AGN population, such as their incidence and power distributions. In particular, X-ray- and radio-selected samples of AGN provide effective means to explore the relevance of radiative and kinetic feedback modes, as they have high purity (less affected by contaminating sources) and completeness (e.g. not affected by dust obscuration). There exist numerous multiwavelength fields which can and have been used for this type of investigation \citep[e.g.][]{Aird2012, Bongiorno2012,Georgakakis2017, Aird2019, KondapallyDeepFields2021, Birchall2022, Kondapally2022}. However, for this work, we focus on the area covered in X-rays by the deep pilot field of the eROSITA All-Sky Survey, called eFEDS \citep{Brunner2022, Teng_eFEDS, Mara_ctp_efeds}; and a co-spatial radio observation taken by the Low Frequency Array (LOFAR) Two-metre Sky Survey \citep[LoTSS;][]{Shimwell2017, Shimwell2019, Pasini2022}. This field is also unique in its richness of multi-wavelength information thanks to the 9hr field of the Galaxy and Mass Assembly (GAMA) DR4 survey \citep[hereafter: GAMA09;][]{Driver2022}, which provides accurate galaxy and environmental parameters to supplement X-ray and radio observations.

This breadth of multi-wavelength data was exploited by \citet{Igo2024}, where we used complete samples to compute the incidence of radio and X-ray AGN from the LOFAR-eFEDS field as a function of various mass-scaled power indicators. A major result of that work was the discovery that, unlike the overall mass-invariant triggering mechanism traced by the X-ray AGN incidence, the radio AGN incidence reveals more complexity, in the form of a residual mass-dependence and a further jet power dependence when considering different radio morphologies \citep[compact versus complex, see \S 3.2.1. in][]{Igo2024}. 

Due to its statistical completeness, the study of incidences as a function of mass-scaled kinetic power provides important information about how AGN with different jet powers and radio morphologies, in galaxies of different masses, release energy into their host galaxies (or halos). For example, the incidence of radio AGN as a function of specific black hole kinetic power, $\lambda_{\rm Jet}$, can inform us about the average jet power of specific subsets of the galaxy population. The same measure of incidence, when convolved with the galaxy stellar mass function \citep{Driver2022, Bernardi2018}, should recover the total radio AGN luminosity function. For example, \citet{Aird2013} use the X-ray AGN incidence previously found in \citet{Aird2012} to compute an X-ray luminosity function in agreement with observations \citep{Aird2010}.

In this work, building on the results of \citet{Igo2024}, we use the same complete sample of low-redshift ($z<0.4$) radio AGN to provide a quantitative measure of the average jet kinetic feedback as a function of galaxy stellar mass ($M_*$), distinguishing between compact and complex radio morphologies. Additionally, we aim to quantify the ability of jet kinetic energy to drive out gas from the host galaxy or halo, as well as its contribution to the heating and cooling balance in the population of massive galaxies (above our completeness limit for the parent sample of host galaxies, $\log (M_*/M_{\odot}) > 10.6$).  This will allow us to estimate, in all generality, the impact of kinetic feedback from radio AGN on different subsets of the galaxy population, ranging from the galaxy ($12<\log (M_h/M_{\odot}) <13$), to the group ($13<\log (M_h/M_{\odot}) <14$) and cluster ($14<\log (M_h/M_{\odot}) <15$) regimes. These estimates are vital for furthering our understanding of black hole accretion processes, including jet powering, as well as black hole-galaxy coevolution and growth.

The paper is organised as follows. Section 2 follows directly from the results of \cite{Igo2024} and details the radio AGN incidence as a function of radio morphology. Section 3 shows how the incidence is used to synthesise the total radio AGN luminosity function. Section 4 describes the methods to calculate average jet power as function of mass and radio morphology, as well as tying this to physical galaxy and halo quantities measuring the efficacy of kinetic feedback. The results found in Sections $2-4$ are then discussed in Section 5. A standard flat cosmology with $H_0 = 70\rm{~km~s^{-1}~Mpc^{-1}}$, $\Omega_M=0.3$, and $\Omega_{\Lambda}=0.7$ is used throughout.

\begin{figure*}[ht!]
\centering
\includegraphics[width=0.497\linewidth]{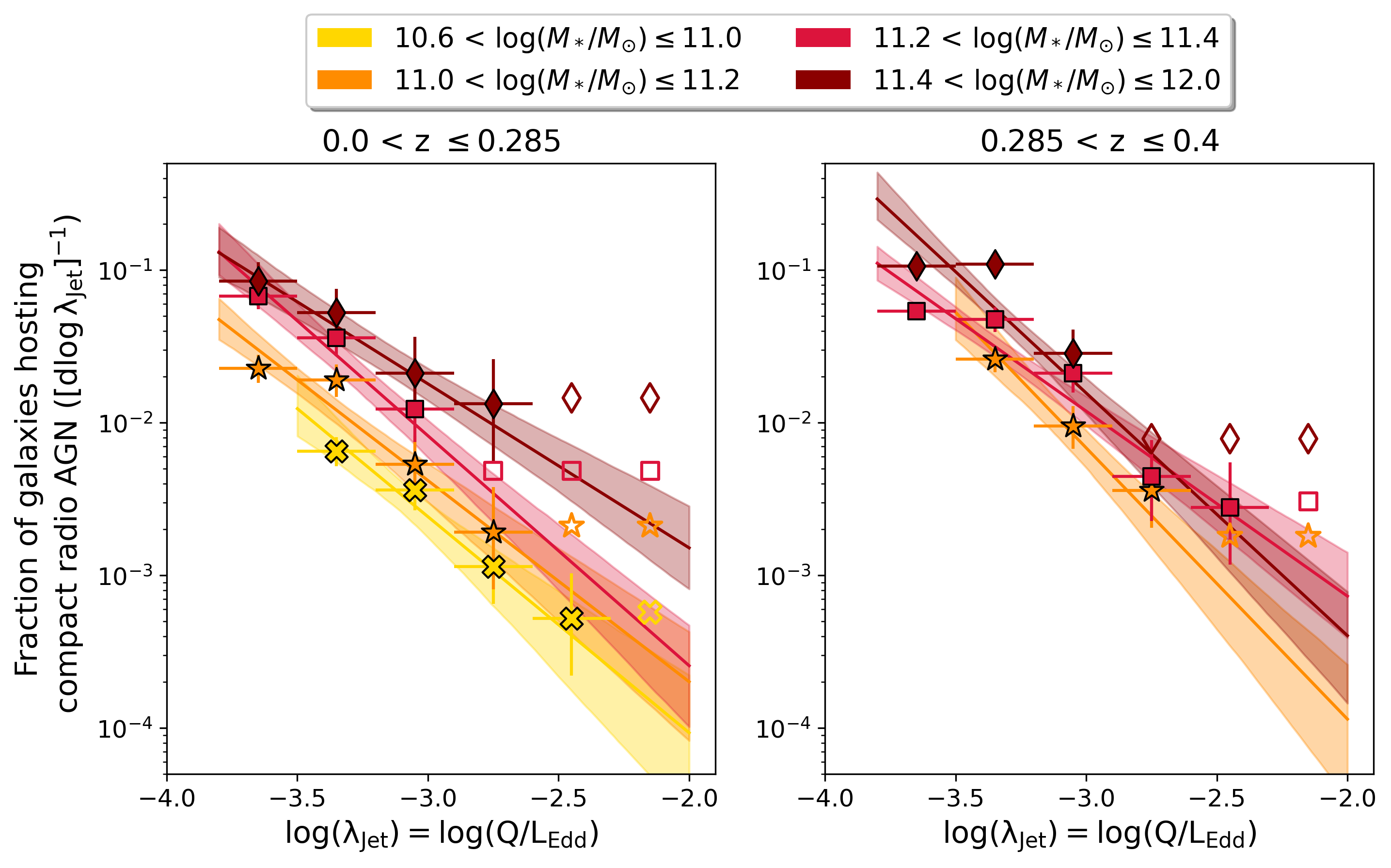}
\includegraphics[width=0.497\linewidth]{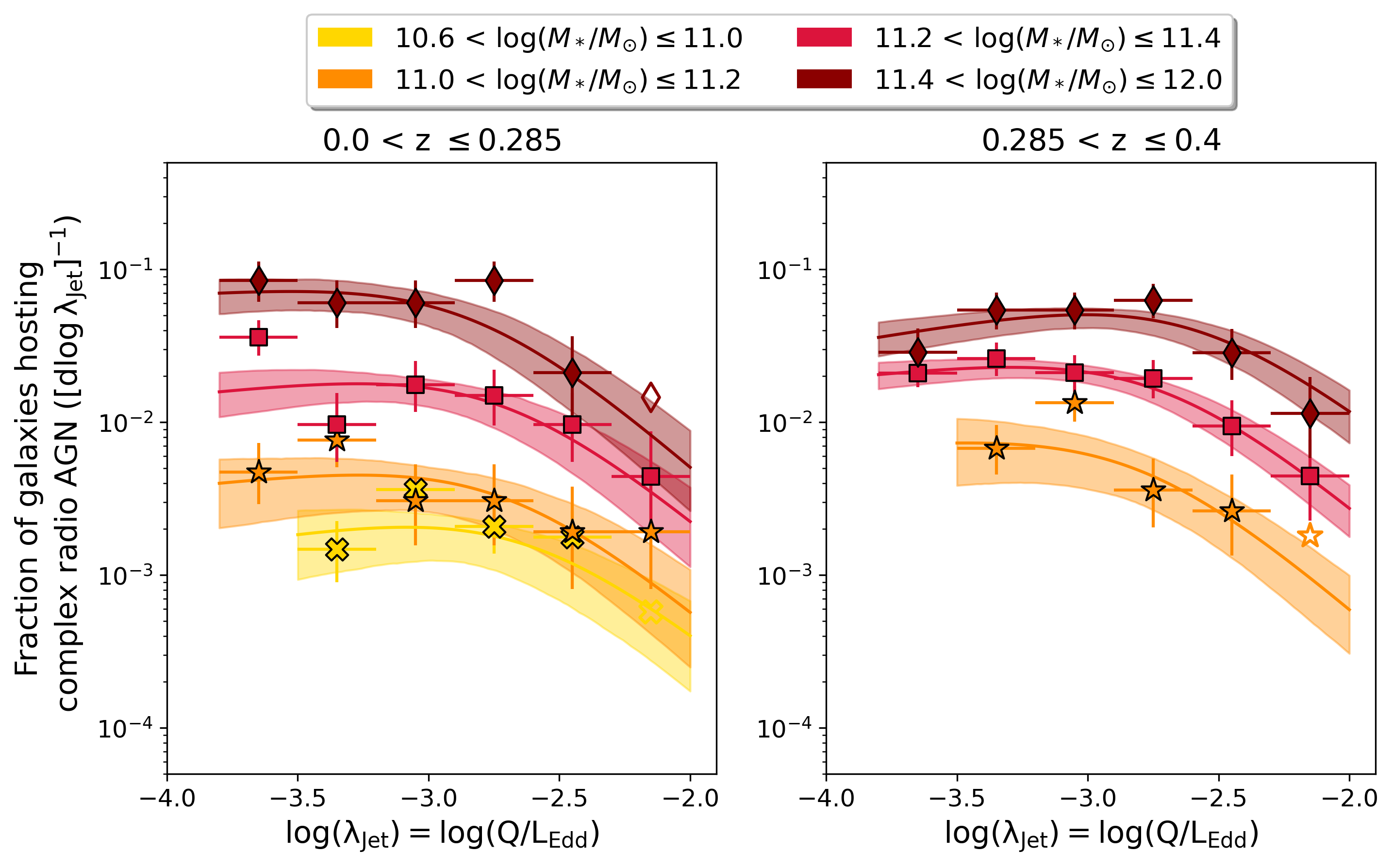}
\caption{Fraction of GAMA09 galaxies hosting compact (leftmost two panels) and complex (rightmost two panels) radio AGN as a function of $\lambda_{\rm{Jet}}$, in different stellar mass and redshift bins, modelled by a single and double power-law, respectively (solid line with 1$\sigma$ error margin). 3$\sigma$ upper limits are shown with unfilled markers.}
\label{fig:compact_complex_incidence}
\end{figure*}

\section{Radio AGN incidence}
\label{sec:incidence_intro}

\citet{Igo2024} presented for the first time an accurate measure of the radio AGN incidence, that is, the fraction of (massive) galaxies hosting radio AGN as a function of the specific black hole kinetic power for compact (see their Figure 19) and both compact and complex radio morphologies (see their Figure 22) in four stellar mass and two redshift bins. For completeness, we briefly summarise the sample selection and methods of \citet{Igo2024} below (and refer the reader to their Section 2, 3, and 4 for more details). 

In terms of the radio morphology classification, there are four criteria that a LOFAR radio source must simultaneously fulfil in order to be considered `compact' \citep[see Section 3.2.1. of][]{Igo2024}: 1) be modelled by a single Gaussian by the PyBDSF algorithm \citep{PyBDSF2015}; 2) have a major axis less than $19.1\arcsec$; 3) have a total to peak flux ratio, $R=F_{\rm Tot}/F_{\rm Peak}$, less than $3.6$; and 4) have no other radio neighbours within 45\arcsec (isolated). If any one of the above criteria is not satisfied, the source is considered a `complex' radio emitter. Optical counterparts from the Legacy Survey are then found for the radio and (separately) X-ray sources in the LOFAR-eFEDS field, using a Bayesian cross-matching algorithm called NWAY \citep{maraNWAY2018}. Matching via optical coordinates to the GAMA09 survey provides accurate stellar masses, star-formation rates and spectroscopic redshifts, which are then used to identify radio and X-ray AGN, by their excess emission compared to other contaminating processes \citep[e.g. star-formation, X-ray binaries; ][]{Mineo2012,Lehmer2016, Best2023radioAGN}. Careful considerations for mass and luminosity completeness, as well as for the purity of the radio AGN sample, are applied to recover robust results. 

The specific black hole kinetic power, $\lambda_{\rm{Jet}}$, is defined as:
\begin{equation}
    \lambda_{\rm Jet}=\frac{Q}{L_{\rm Edd}},
    \label{eq_jetpower_Ledd}
\end{equation}
where the $L_{\rm Edd}$ is the Eddington luminosity, estimated here using a linear scaling between black hole mass and stellar mass $M_{\rm BH}=0.002M_*$ \citep{Marconi2003} and $Q$, the jet power, is defined as \citep[][their Eq.~2]{HeckmanandBest2014}:
\begin{equation}
    Q = 2.8 \times 10^{37}~ \Bigg(\frac{L_{\rm 1.4GHz}}{10^{25}~{\rm W~Hz}^{-1}}\Bigg)^{0.68}~{\rm W},
    \label{Qeq}
\end{equation}
a scaling relation derived from observed radio jets creating X-ray cavities in nearby clusters. The LOFAR 144MHz radio luminosity is converted to 1.4~GHz radio luminosity assuming a spectral index of $\alpha=0.7$, using the following convention for the radio flux density ($S_\nu$) as a function of frequency ($\nu$): $S_\nu \propto \nu^{-\alpha}$.

Since one of the aims of this work is to quantitatively understand how radio morphology impacts the power, efficacy and scale of the kinetic feedback, we re-compute the incidence individually for compact and complex radio AGN. It is important to highlight here that the split in radio morphology is dependent on the resolution of the survey, in this case the 8\arcsec\ $\times$ 9\arcsec\ resolution of the LOFAR-eFEDS field, so small-scale sources may become resolved at higher resolution. Interestingly, LOFAR Very Long Baseline (LOFAR VLBI) studies find that small-scale sources may become resolved complex sources at higher resolution (0.3\arcsec), although only 40\% of unresolved sources detected at the Dutch station resolution of 6\arcsec\ have a high resolution counterpart, and of this sample 89\% remain compact \citep[][]{Morabito2022, Sweijen2022}. The radio AGN incidence is computed in bins of $\Delta \log \lambda_{\rm{Jet}}=0.3$, where the incidence is defined as the probability density per logarithmic $\lambda_{\rm{Jet}}$ interval (units of $[\log \lambda_{\rm{Jet}}]^{-1}$).

Figure \ref{fig:compact_complex_incidence} (left) shows the compact radio AGN incidence, fit by a simple power-law in the form of:
\begin{equation}
    f(x)=A~\Bigg(\frac{x}{x_0}\Bigg)^B,
    \label{eq:pow}
\end{equation}
where $A$ is the normalisation, $\log(x_0)=-3.25$ and $B$ is the slope. Bayesian fitting, using UltraNest\footnote{\url{https://johannesbuchner.github.io/UltraNest/example-line.html}} \citep{UltraNest2021} is performed by sampling an asymmetric point cloud defined by the 1$\sigma$ uncertainties on the values \citep[][upper limits are sampled with a bounded box at the 3$\sigma$ value]{Cameron2011}. Results of the power-law fitting are shown in Fig.~\ref{fig:compact_incidence_fit_results}, where a clear increase in normalisation, i.e. a mass-dependence, is seen in both low (green) and high (purple) redshift bins (left panel), whilst the slope remains relatively constant around $-1.4\pm0.2$ (right panel), as already discussed in \citet{Igo2024}. 

Figure \ref{fig:compact_complex_incidence} (right) instead shows the complex radio AGN incidence, fit by a double power-law described by Eq.~\ref{eq:doubpow},
\begin{equation}
    f(x) = \frac{k}{\left(\frac{x}{x_b}\right)^{-\alpha_1}+\left(\frac{x}{x_b}\right)^{-\alpha_2}},
    \label{eq:doubpow}
\end{equation}
where $k$ is the normalisation, $x_b$ is the break and $\alpha_1$, $\alpha_2$ are the faint-end and bright-end slopes, respectively. The power-law indices are well-fit on average by a shallow faint-end, $\alpha_1\sim0.2$, and a steep bright-end, $\alpha_2\sim-1.4$, slope. Similarly to the compact-only fits, other than the normalisation of the incidence ($k$), the other fit parameters remain constant within the standard error deviation, as shown in Fig.~\ref{fig:complex_incidence_fit_results}. We note that the double power-law fit provides an adequate parameterisation of the incidence distribution for the purpose of this work, but that future larger samples could highlight more complexity in the shape of the distribution, especially when looking into the redshift-evolution in more detail.

Fig.~\ref{fig:compact_complex_incidence} highlights explicitly the stark differences present between the $\lambda_{\rm{Jet}}$ distributions of compact versus complex sources, already found in \citet{Igo2024} (see their Figure 22, where a different, but consistent, functional form was used to fit the incidence curves)\footnote{Note that the results presented in this section are fully consistent with \citet{Igo2024} and are independent of the choice of binning, as shown in Fig.~\ref{fig:consistent_past_incidence}, plotting the incidence of both compact and complex radio AGN.}. 
Overall, the steep power-law-like incidence of compact radio AGN (Fig.~\ref{fig:compact_complex_incidence}, left) indicates that compact (i.e. unresolved by LOFAR) radio AGN dominate the lower jet powers and drop out rapidly at higher jet powers. In contrast, the flatter incidence distribution of complex radio AGN, and their stronger mass scaling (Fig.~\ref{fig:compact_complex_incidence}, right) indicates that complex radio AGN progressively dominate at high $M_*$ and $\lambda_{\rm Jet}$. Interestingly, the break in the jet power distribution occurs around the point where the fraction of Fanaroff-Riley II \citep[FRII;][]{Fanaroff&Riley1974} double-lobed radio AGN out of the total compact and complex radio AGN sample increases rapidly. The average secure FRII fraction, defined as those classified unanimously by visual inspection to have a double-lobed, edge-brightened structure \citep[see Appendix C of][]{Igo2024}, between $-3.8<\log \lambda_{\rm Jet}\leq-2.9$ is 4.6\%, whereas between $-2.9<\log \lambda_{\rm Jet}<-2.0$ it reaches 37.3\% (averaged across all mass and redshift bins). A larger sample would be needed to understand the true nature of this break, as it could be an indication of a change of accretion mode \citep[see \S 6.3 of][]{Igo2024}, inadequacy of a universal $Q-L_R$ relation (see \S \ref{discussion:caveatsQ}) or something else entirely.

In the following sections we will elaborate on the implications of such stark differences in the radio AGN incidence and jet power distributions for the synthesis of the radio AGN luminosity function, and for our estimates of AGN kinetic feedback for galaxies (and halos) of different masses.

\section{Synthesis of the radio luminosity function}
\label{sec:LFs}

Under the assumption that our samples are complete (or completeness-corrected), convolving the incidence distributions --- the functions describing how many galaxies host radio AGN in a given luminosity range --- with the galaxy stellar mass function (SMF) --- a measure of how many galaxies there are in a given stellar mass range per comoving volume --- is an alternative way to recover the AGN radio luminosity function (RLF). The RLF describes how many radio AGN there are within a given luminosity range per comoving volume.

This is analogous to previous works, for example \citet{Aird2013}, where it was shown to be possible to recover the X-ray luminosity function (XLF) by convolving the X-ray AGN incidence, as computed in \citet{Aird2012}, with the SMF from \citet{Moustakas2013}. Based on our work in the eFEDS/GAMA09 field, we could also reproduce the results of \citet{Aird2013}: the eFEDS X-ray AGN incidence results presented in \citet{Igo2024} are indeed in good agreement with \citet{Aird2012}; therefore, we do not discuss XLFs further in this work. Instead, we focus here on computing the RLF and its dependence on radio morphology and host stellar mass. 

Following the method of \citet{Aird2013}, the RLF of AGN can be expressed as:
\begin{eqnarray}
\psi(L_\mathrm{R}, z) & =  & \phi(M_* , z) * p(\lambda_{\rm {Jet}} \;|\; M_*,z) \nonumber \\
				   & = & \int \phi(M_* , z) \; p\big(\lambda_{\rm {Jet}} \;|\; M_*,z\big) \frac{\;\mathrm{d} \log(\lambda_{\rm {Jet}})}{\;\mathrm{d} \log(L_R)} \;\mathrm{d} \log M_* 
\label{eq_convolution}
\end{eqnarray}
where $\phi(M_*, z)$ is the galaxy SMF and $p(\lambda_{\rm {Jet}} \;|\; M_*,z)$ describes the probability in $[\log \lambda_{\rm{Jet}}]^{-1}$ units, for a galaxy of given $M_*$ and $z$ to host an AGN with a specific kinetic power, $\lambda_{\rm{Jet}}$. 

\begin{figure}[t!]
\centering
\includegraphics[width=\linewidth]{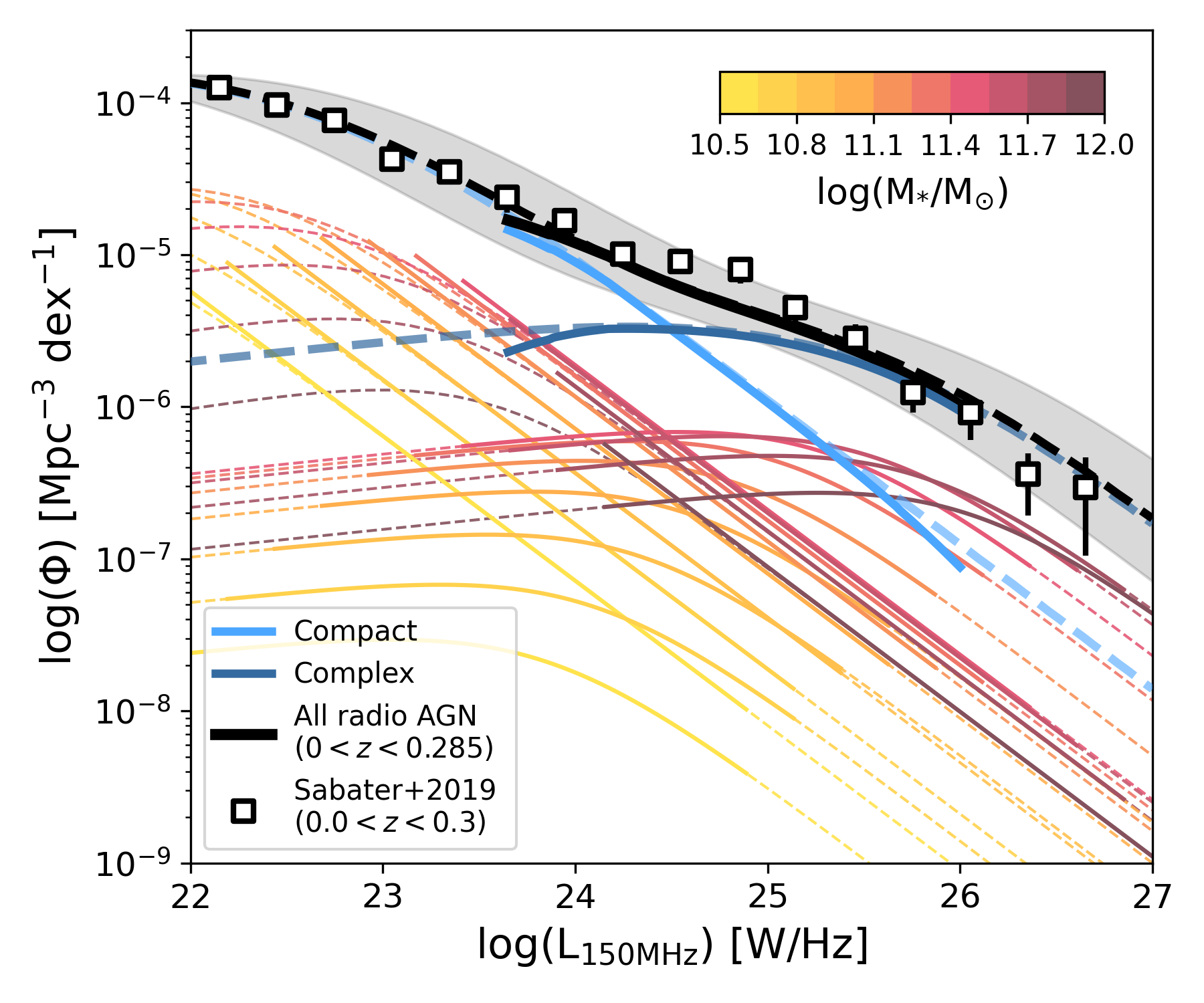}
\caption{Total radio luminosity function (RLF; black solid line), derived by convolving the radio AGN incidence with the stellar mass function, decomposed into the contribution from radio AGN of varying radio morphology (compact -- light blue; complex -- dark blue) and host galaxy stellar mass (same colourmap as Fig.~\ref{fig:compact_complex_incidence}). The RLF is in good agreement with \citet{Sabater2019} (squares; error bars often too small to be seen). Dashed lines indicate extrapolation of the incidence distribution, and thus the RLF, and the grey shaded region represents the uncertainty on the RLF (see text for more details).}
\label{fig:RLF}
\end{figure}

Given the use of GAMA09 data in this work, the appropriate SMF would be the one using GAMA galaxies from \citet{Driver2022}. However, as seen in their Figure 12 and as pointed out by the authors, the GAMA SMF is unable to accurately capture the high-mass end due to the small survey volumes. For our study on radio AGN, which is complete only for massive galaxies ($\log (M_*/M_{\odot}) > 10.6$), we therefore adopt the SMF from \citet{Bernardi2018} who use the larger volume Sloan Digital Sky Survey (SDSS) and find consistent results with \citet{Driver2022} up to $\log (M_*/M_{\odot}) \sim 11$.

Figure \ref{fig:RLF} shows our derived RLF (black), in the redshift range $0<z<0.285$, broken down, for the first time, into the contribution from compact (light blue) and complex (dark blue) radio AGN and their respective contributions in different host stellar mass bins (same colourbar as Figure \ref{fig:compact_complex_incidence})\footnote{Note that we only plot the range of massive galaxies with $10.5<\log(M_*/M_{\odot})<12.0$, meaning that at the lowest luminosities, where lower mass galaxies dominate, the total RLF shown in Fig.~\ref{fig:RLF} should be regarded as a lower-limit.}.

The RLF is plotted with a solid line in the range $\log(L_R/[\rm{W~Hz^{-1}}])= 23.65-26$, where the lower bound corresponds to the 95\% radio luminosity completeness level at $z=0.285$ \citep[see Figure 11 in][]{Igo2024} and the upper bound is driven by the LOFAR-eFEDS survey's volume-limit. To obtain the RLF in this range, only the radio AGN incidence distributions in the observed range of $-4 <\log \lambda_{\rm {Jet}}<-2$ (see Fig.~\ref{fig:compact_complex_incidence}) are convolved with the SMF as described in Eq.~\ref{eq_convolution}. 

\begin{figure}[t!]
\centering
\includegraphics[width=\linewidth]{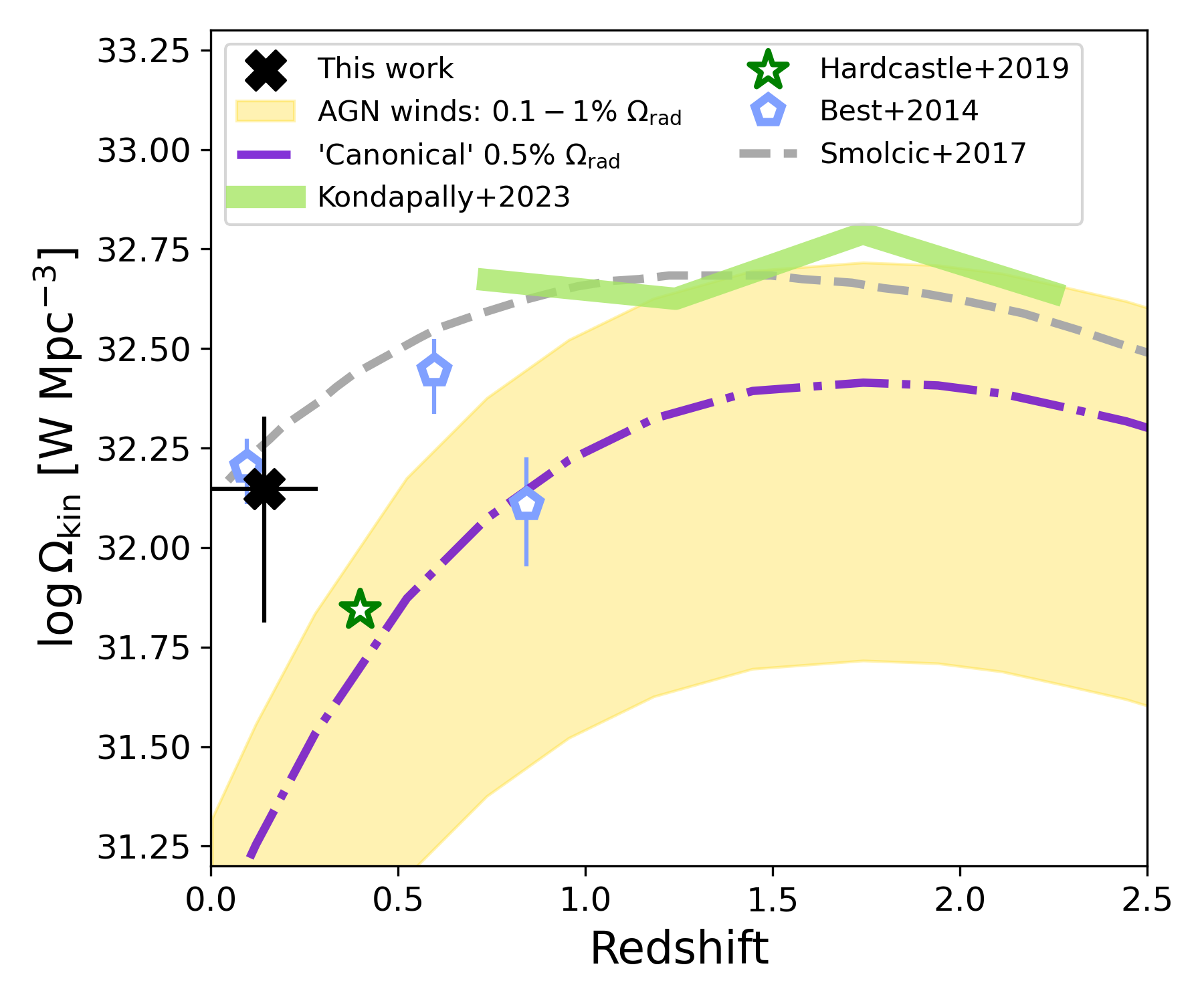}
\caption{Kinetic luminosity density of (radio) AGN as a function of redshift obtained from recent observational work. Results of this work are shown by the thick black cross. The yellow shaded region also shows the $0.1-1$\% of bolometric luminosity density from \citet{Aird2015} which represents a rough indicator of the `canonical' feedback energy that may be delivered by radiatively-driven AGN winds.}
\label{fig:kond2023}
\end{figure}

The dashed lines in Fig.~\ref{fig:RLF} are extrapolations to the incidence distributions, given the assumptions that follow, in order to sample the faint and bright ends of the RLF. Firstly, for the incidence of complex radio AGN, we simply extrapolate the low- and high-ends of the double power-law in the range $-8 <\log \lambda_{\rm {Jet}}<0$ and perform the same convolution procedure as described in Eq.~\ref{eq_convolution}. We do a similar extrapolation for the incidence of compact radio AGN, except for the low-power end we assume a mass-independent turnover at $\log \lambda_{\rm {Jet}}=-4.4$ after which a flatter power-law with slope $0.3$ is followed. With the turnover being beyond the parameter space covered in this work, these values are chosen such that the extrapolated RLF agrees best with literature results (see \S \ref{discussion:rlf} for more discussion). As low power compact sources are very numerous (the incidence function is steep), this turnover is important: the incidence is a probability distribution, therefore it must be bounded at its extremes and must integrate to one, conditions that a simple power-law cannot satisfy. We apply the same extrapolation to the data in the $0.285<z<0.4$ redshift bin.

The grey shaded region in Fig.~\ref{fig:RLF} represents the uncertainty on the RLF. Given the method to obtain the RLF in this work via Eq.~\ref{eq_convolution}, the major sources of uncertainty reside in the accuracy of: the SMF, the modelling of incidence distributions, the stellar mass and jet power determination. Having already motivated the choice of SMF for this work above, carefully modelled the incidence of compact and complex radio AGN in Section \ref{sec:incidence_intro} and justified the relatively low ($\sim 0.1$~dex) uncertainty in the stellar mass \citep[see Fig. 25 in][]{Igo2024}, we deem the major uncertainty in the RLF originates from the determination of jet power $Q$ itself \citep[see detailed discussion in \S 6.2.1 of][]{Igo2024}. We account for this by adding a $\pm0.3$~dex scatter on the jet power but keeping all aforementioned extrapolation parameters the same. The choice of $\pm0.3$~dex is motivated by the scatter (over the range of jet powers probed by this work) in the inferred jet power distribution presented in \citet{Hardcastle2019}, which takes into account relevant jet physical processes and source environments.

Given the assumptions above, we reproduce well the full range of the RLF, in line with the results of \citet{Sabater2019} shown by square markers on Fig.~\ref{fig:RLF} \citep[see also][for more comparison with recent literature RLFs]{Kondapally2022}. Overall, we find that compact (complex) radio AGN dominate in number density below (above) $\log (L_{\rm{150MHz}}/[{\rm W~ Hz}^{-1}]) \sim 24.5 $, equivalent to a jet power of $\log (Q/[\rm W]) \sim 36.7$. Additionally, the number density of radio AGN with typical luminosities $\log (L_{\rm{150MHz}}/[{\rm W~ Hz}^{-1}]) \sim 23.5-25 $ is dominated by galaxies of mass $\log (M_*/M_{\odot}) \sim 11.4$, with broad $\log \lambda_{\rm Jet}$ distribution ($\sim -3.8$ to $-2.6$). Given that radio AGN preferentially lie in higher mass galaxies, the predominant contribution by moderately massive galaxies to the RLF is \textit{not} just due to the shape of the SMF (which has a break around $\log (M_*/M_{\odot}) \sim 11.4$), unlike the mass-independent X-ray AGN incidence and the XLF results by \citet{Aird2013}. Above $\log (L_{\rm{150MHz}}/[{\rm W~ Hz}^{-1}]) \sim 25 $, radio AGN with host $\log (M_*/M_{\odot}) > 11.4$ and $\log \lambda_{\rm Jet}$ reaching up to $-1.7$, dominate in number density, whereas below $\log (L_{\rm{150MHz}}/[{\rm W~ Hz}^{-1}]) \sim 23.5$, the main contributors are radio AGN in galaxies with $\log (M_*/M_{\odot}) < 11.4$ and reaching down to $\log \lambda_{\rm Jet} \sim -4.1$. 

By transforming the observed radio luminosity into a jet power using Eq.~\ref{Qeq}, i.e. computing a jet kinetic luminosity function, $\rho(Q, z)$, and integrating as in Eq.~\ref{eq:omegakin}, we also derive the average jet kinetic heating rate per Mpc$^{-3}$, $\Omega_{\rm kin}$, in other words the kinetic luminosity density of the whole AGN population.
\begin{equation}
\label{eq:omegakin}
\Omega_{\rm{kin}}(z) = \int Q ~ \rho(Q,z) ~ \mathrm{d} Q ~\rm{W\,Mpc^{-3}}.
\end{equation}
Propagating the uncertainties from the grey shaded region on Fig.~\ref{fig:RLF}, we obtain $\log \Omega_{\rm kin}/[\rm {W~Mpc^{-3}}]=32.15_{-0.34}^{+0.18}$ in the redshift range $0<z<0.285$, which is in a similar range to past observational works as shown in Figure \ref{fig:kond2023} \citep{Best2014,Smolcic2017, Hardcastle2019, Kondapally2023}.

Beside the kinetic luminosity density, we can roughly estimate the total radiative wind-driven outflow energy density by scaling the (well-constrained) AGN bolometric luminosity density as a function of redshift \citep[see e.g.][]{MerloniHeinz2008, Aird2015} by a factor $0.1-1$\%. We show this average quasar-driven wind energy density as a yellow shaded region in Fig.~\ref{fig:kond2023}. This range indicates the typical expectation, observationally, for the fraction of AGN radiative output that can efficiently couple to the surrounding medium (henceforth, `coupling efficiency'), producing feedback effects from multi-phase winds. However, this `coupling efficiency' is widely debated and can span orders of magnitude from $\sim0.001\%$ to tens of percent of the AGN bolometric luminosity ($L_{\rm bol}$), as discussed in detail in \citet{Harrison2018, HeckmanBest2023}, with observational and simulation-based estimates tending to lower and higher values, respectively\footnote{A large scatter in the `coupling efficiency' is thought to result from numerous factors such as observational limitations in accurately measuring outflow geometries and/or energetics, variability in AGN luminosity and/or outflow power, varied sub-grid physics implementation and resolution of different feedback simulations, uncertainty in the fraction of feedback energy in kinetic form, among many others \citep[see e.g.][and references therein for more details]{Costa2014, Harrison2018, Weinberger2018, Ward2024, Harrison2024}.}.
For example, the seminal work of \citet{DiMatteo2005} using hydrodynamical simulations \citep{Springel2005} of galaxy mergers find that their results can reproduce the $M-\sigma$ relation \citep{Ferrarese&Merritt2000} given a `coupling efficiency' of 5\% $L_{\rm bol}$. On the other hand, \citet{HeckmanBest2023} review a wide range of recent observational results spanning the molecular, warm-ionized and highly-ionized phases of radiatively-driven feedback in a variety of AGN, finding a total `coupling efficiency' of 0.5\% $L_{\rm bol}$. Therefore, throughout this work, we assume 0.5\% $L_{\rm bol}$ as the `canonical' fraction of radiative output from multi-phase winds (shown with a purple dot-dashed line in Fig.~\ref{fig:kond2023}) and adopt this reference value when (qualitatively) comparing radiative versus kinetic energetics results.

The global estimates presented in Fig.~\ref{fig:kond2023} show that kinetic feedback from (jetted) AGN dominates over any plausible inventory of radiatively-driven feedback modes. This holds for galaxies with mass comparable to that of the Milky Way and above, at low redshift (see \S \ref{sec:discussion} for more discussion). In the following section we proceed to analyse how this dominant feedback mode distributes over systems (galaxies, halos) of different mass, providing fresh insights into the mechanics of AGN feedback in the large scale structure.

\section{Global energetics of radio AGN kinetic feedback}
\label{sec:energetics}

Having shown that the RLF computed via the incidence distributions agrees with the observed one, in this section we proceed to compute reliable estimates for the global energetics of radio AGN kinetic feedback. 

\subsection{Average jet power of massive galaxies}

The average jet power, $\overline{Q}(M_*)$, released by massive galaxies is obtained by computing the expectation value of the modelled incidence distribution functions, shown in Fig.~\ref{fig:compact_complex_incidence}, as a function of stellar mass. For the compact radio AGN case, we take the average value ($-1.4$) of the power-law slope of the incidence function across the two redshift bins and also determine the power-law normalisation, given by the best-fit (and intrinsic scatter) to the observed incidence distributions in each redshift bin as a function of stellar mass (see Fig.~\ref{fig:compact_incidence_fit_results}, left). The observed distributions (Fig.~\ref{fig:compact_complex_incidence}) are constrained by our data only over a limited range in $\lambda_{\rm{Jet}}$; thus we proceed with the extrapolation as described in Section \ref{sec:LFs} in the range $-8 <\log \lambda_{\rm {Jet}}<0$. For the complex radio AGN case, we take the average values of all double power-law parameters except for the normalisation, which we similarly determine from the best-fit as a function of stellar mass (see Fig.~\ref{fig:complex_incidence_fit_results}) and extrapolate as described in Section \ref{sec:LFs}. The combined weighted average of $\overline{Q}(M_*)$ in each of the two redshift bins is the quantity plotted on Fig.~\ref{fig:compact_complex_incidence}, indicating the average jet power as a function of mass in the full redshift range, $0<z<0.4$.

\begin{figure}[t!]
\centering
\includegraphics[width=\linewidth]{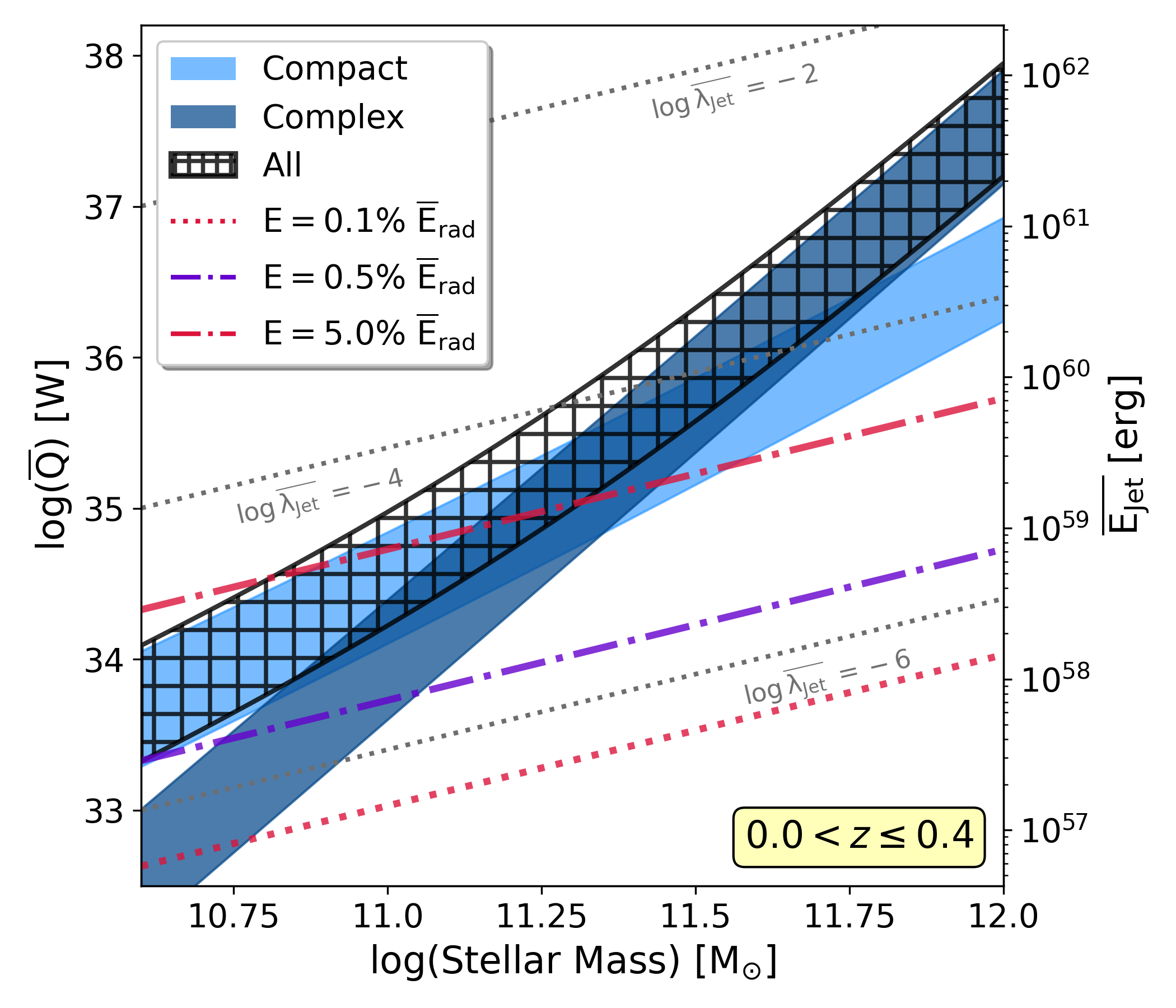}
\caption{Average jet power released by radio AGN as a function of stellar mass for the population of all massive galaxies between $0<z<0.4$ (black hatched). The secondary y-axis (right) shows the corresponding average jet kinetic energy released, obtained by multiplying $\overline{Q}$ with the look-back time. The contributions from compact and complex radio AGN are shown with light and dark blue shaded regions, respectively. Red dotted and dot-dashed lines indicate the 0.1\% and 5\% of the average radiative energy output as a function of $M_*$, respectively. The purple dot-dashed line represents a `canonical' 0.5\% of the average radiative energy output, which is used for comparison to the kinetic output throughout the paper.}
\label{fig:energetics_Qav}
\end{figure}

Figure \ref{fig:energetics_Qav} shows the average jet power output of massive galaxies, considering LOFAR-detected and undetected sources (via the incidence), as a function of $M_*$. The intrinsic scatter on the best-fit normalisation as a function of stellar mass (Figs~\ref{fig:compact_incidence_fit_results} and \ref{fig:complex_incidence_fit_results}) is propagated through the calculation of $\overline{Q}$ to find the lower and upper uncertainty on this quantity, shown by the shaded regions. Light and dark blue shaded regions mark the compact and complex contributions, respectively, with the black hatched region defining their sum, considering all massive galaxies.  $\overline{Q}$ is also converted to an average jet kinetic energy released, $\overline{E_{\rm{Jet}}}$ (rightmost y-axis), with a multiplication by the look-back time between $0.0<z<0.4$. 
As a comparison, 0.1\% (red dotted), 0.5\% (purple dot-dashed) and 5\% \citep[red dot-dashed;][]{DiMatteo2005} of the average radiative energy output as a function of $M_*$ ($\overline{E_{\rm{rad}}}$) are also shown\footnote{$\overline{E_{\rm{rad}}}$ is computed by firstly integrating the specific black hole radiative power distribution, $p(\lambda_{\rm {Edd}} \;|\; M_*,z)$, to find the average $\lambda_{\rm {Edd}}$, using Model C of \citet{Aird2013} at the average redshift ($z=0.27$) of the radio AGN sample. Then the corresponding bolometric luminosity as a function of $M_*$ is converted to $\overline{E_{\rm{rad}}}$, multiplying by the look-back time to $z=0.4$.}.

Figure \ref{fig:energetics_Qav} highlights that the average jet power of a population of massive galaxies in the local universe ranges from $\log (\overline{Q}/[{\rm W}]) \sim 33.7-37.5$ and is dominated by compact (complex) radio AGN at stellar masses below (above) $\log M_*/M_{\odot} \sim 11.5$. More specifically, the total kinetic energy released by complex radio sources grows more steeply with stellar mass than that released by compact sources, reflecting the different $M_*$ dependence of the incidence normalisation. The average $\lambda_{\rm Jet}$ is $\sim 10^{-5}-10^{-3}$ across the mass range of this study. 

When comparing to the radiative energy output of AGN, we see that the average jet power of the entire population also increasingly dominates over the `canonical' fraction ($0.5\%$) of average radiative power useful for feedback as stellar mass increases. Due to the observed near mass-independence of the X-ray AGN incidence \citep{Aird2012}, the radiative output as a function of $M_*$ has a shallower (linear) slope than the kinetic output released by both compact and complex radio AGN. For the assumed `canonical' value of $0.5\%~L_{\rm bol}$, the radiative mode of feedback only begins to dominate at low stellar masses, $\log M_*/M_{\odot} < 10.6$. 

\subsection{Disruptive kinetic feedback in massive galaxies and dark matter halos}

We are now in the position to compare $\overline{E_{\rm{Jet}}}$ to the binding energy, $U_{\rm bin}$, of the host galaxy (small scales) and host dark matter halo (on large scales). A total integrated kinetic energy injection in excess of a galaxy/halo binding energy would suggest that jetted radio AGN may have the ability to disrupt the stellar body and gas distribution of their host galaxies or halos, or at least deeply influence it. Therefore, we define this ratio as the `small-scale (or large-scale) disruptive kinetic feedback efficiency', $\mathcal{F}_{\rm D,~ gal/halo}=\overline{E_{\rm{Jet}}}/U_{\rm bin,~gal/halo}$. 

In order to estimate the galaxies' binding energy, we use the result of \citet{Shi2021}, who compile a sample of 752 objects from the literature, ranging over a wide variety of galaxy types, that have measurements for their $M_*$, effective radius\footnote{The effective radius of a galaxy is defined as the radius within which half of the total light is emitted.} ($R_{\rm e}$) and the dynamical velocity ($V_{\rm e}$) at
$R_{\rm e}$. They find a strong empirical correlation between $V_{\rm e}R_{\rm e}^{0.25}$ ---effectively the fourth-root of galaxy binding energy, $U_{\rm bin, gal}$--- and the galaxy stellar mass:
\begin{equation}
\label{eqn_Ubin_gal}
  \begin{split}
    U_{\rm bin, gal} &\approx \frac{GM_{\rm dyn, e}^{2}}{R_{\rm e}} \equiv \frac{R_{\rm        e}V_{\rm e}^{4}}{G} \\
    &=4.62\times10^{48} \;{\rm erg} ~ \left( \frac{V_{\rm e}R_{\rm        e}^{0.25}}{\rm km\;s^{-1}\;kpc^{0.25}} \right)^{4},
  \end{split}
\end{equation}
where $M_{\rm dyn, e}$ is the dynamical mass within $R_{\rm e}$ and 
\begin{equation}
V_{\rm       e}R_{\rm        e}^{0.25} =  15.7~\mathrm{km~s}^{-1}~\mathrm{kpc}^{0.25} \left( \frac{M_{*}}{M_{0}} \right)^{0.134} \left( 1+ \frac{M_{*}}{M_{0}} \right)^{0.272},
\end{equation}
with $M_{0}=2.5\times10^{7}$\;M$_{\odot}$.

For halos, we use simple physical arguments to obtain the binding energy of the baryons in the halo, as follows:

\begin{equation}
    \label{eqn_Ubin_halo}
    U_{\rm bin, halo} = \frac{3}{5} ~ \frac{G ~f_{\mathrm{gas}} M_h^2}{R_{200c}},
\end{equation}
where the radius, $R_{200c}$, at 200 times the critical density of the universe at a given redshift, $\rho_{200c}(z)$, is:
\begin{equation}
    R_{200c}=\frac{3}{4 \pi} ~ \Bigg(\frac{M_h}{\rho_{200c}(z)}\Bigg)^{\frac{1}{3}}.
\end{equation}
Note that the gas fraction, $f_{\rm{gas}}$ is the ratio of the cosmological baryon and total matter density ($\Omega_b/\Omega_M$) and its universal value is $\sim 0.16$ \citep[e.g. see review of][and references therein]{Eckert2021}. 

\begin{figure}[t!]
\centering
\includegraphics[width=\linewidth]{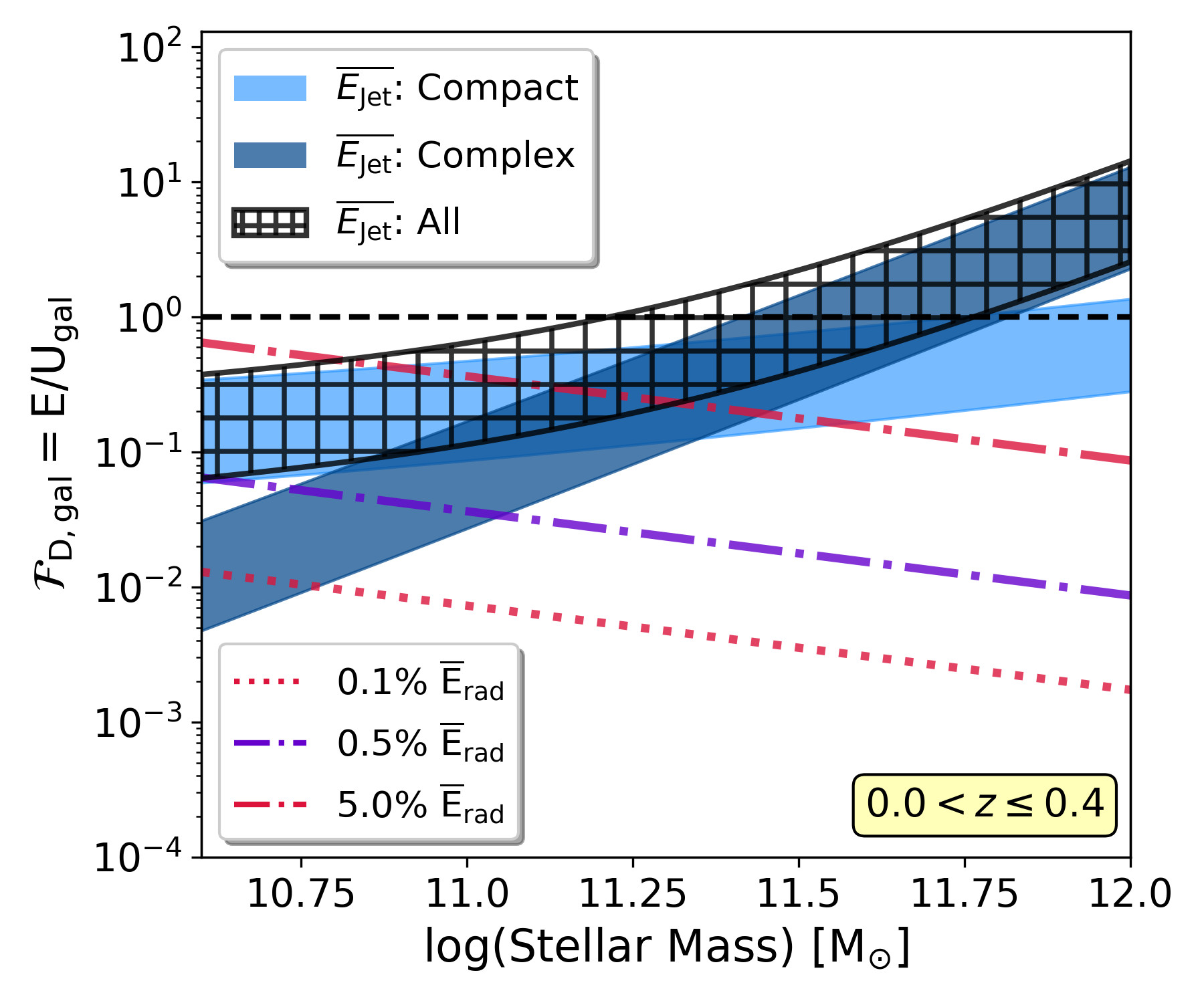}
\caption{Small-scale disruptive kinetic feedback efficiency (i.e. ratio of jet kinetic energy divided by host galaxy binding energy) as a function of stellar mass for the massive galaxy population (black hatched) in the redshift range $0<z<0.4$. The respective contributions from compact and complex radio AGN are shown in light and dark blue shaded regions, respectively. The red and purple lines are the same as in Figure \ref{fig:energetics_Qav}. The horizontal black dashed line marks the equality of the jetted kinetic energy and the host galaxy binding energy.}
\label{fig:energetics_gal}
\end{figure}

\begin{figure}[t!]
\centering
\includegraphics[width=\linewidth]{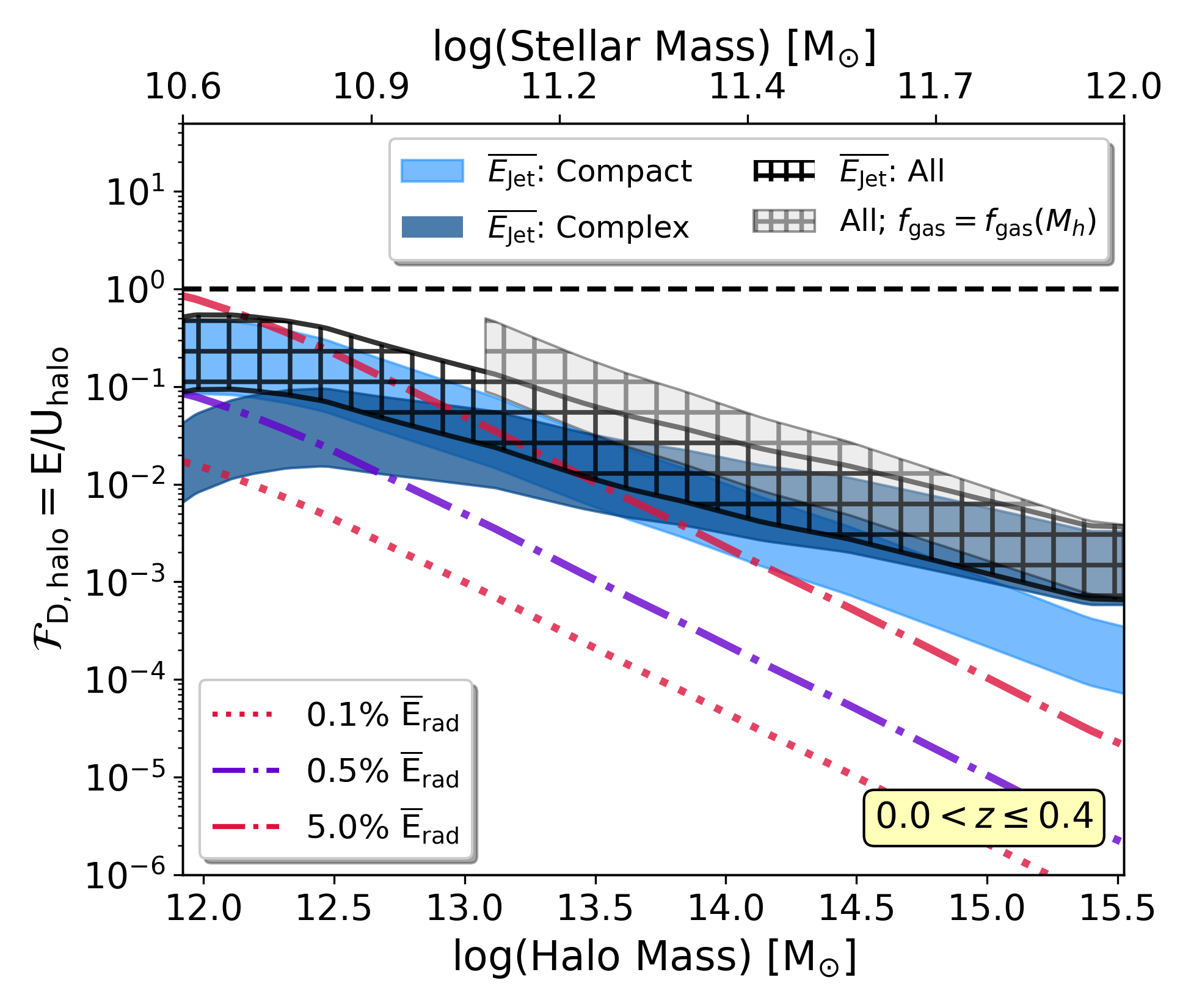}
\caption{Large-scale disruptive kinetic feedback efficiency (i.e. ratio of jet kinetic energy divided by host halo binding energy) as a function of halo mass for the massive galaxy population (black hatched) in the redshift range $0<z<0.4$. The respective contributions from compact and complex radio AGN are shown in light and dark blue shaded regions respectively. The grey-shaded area shows the large-scale disruptive feedback efficiency on the baryonic halo mass (assuming a mass-dependent gas fraction). The red and purple lines are the same as in Figure \ref{fig:energetics_Qav}. The horizontal black dashed line marks the equality of the jetted kinetic energy and the host halo binding energy.}
\label{fig:energetics_halo}
\end{figure}

Figures~\ref{fig:energetics_gal} and \ref{fig:energetics_halo} show the small-scale and large-scale disruptive kinetic feedback efficiency as a function of stellar mass and halo mass for the massive galaxy population (black hatched), with respective contributions from compact (light blue) and complex (dark blue) radio AGN. At the horizontal black dashed line the jetted kinetic energy equals the host galaxy or halo binding energy. In Figure \ref{fig:energetics_halo}, we use the simulation results from \citet{Girelli2020} (their Eq. 6 with best-fit parameters from Table 1) to produce a mapping between $M_*$ and $M_h$ (the so-called Stellar to Halo Mass Relation, SHMR), where $M_h=M_{200c}$.

Thusfar, we have assumed a universal hot gas fraction, however, it is known that $f_{\rm{gas}}(M_h)$ is not constant: groups are more hot gas depleted compared the universal value, typically observed in clusters. To account for this, we adjust Eq.~\ref{eqn_Ubin_halo} using the best-fit relation\footnote{Note that the $f_{\rm gas}-M_h$ relation uses $M_{500c}$ (total mass within $r_{500c}$), so we use Eq. 6 and A2 from \citet{Bocquet2016} to convert $M_{200c}$ from \citet{Girelli2020} to the required $M_{500c}$.} from \citet[][Eq. 11]{Eckert2021}:
\begin{equation}
	f_{\rm gas,500} = 0.079_{-0.025}^{+0.026} \times \left(\frac{M_{500c}}{10^{14}M_\odot}\right)^{0.22_{-0.04}^{+0.06}}.
\end{equation}
We show the resulting $\mathcal{F}_D-M_h$ trend in grey hatched on Fig.~\ref{fig:energetics_halo}, only for $\log M_h/M_{\odot}>13$ where observational measurements for the gas fractions exist.

Upon computing the small-scale (galaxy-wide) disruptive kinetic feedback efficiency, shown in Figure \ref{fig:energetics_gal}, we observe that compact radio AGN have significant $\mathcal{F}_{\rm D,~ gal}$, ranging from $20-80\%$, whereas complex radio AGN have negligible feedback efficiency ($\mathcal{F}_{\rm D,~ gal}\approx 0.02$) for Milky-Way-like galaxies, but reach significant values for massive galaxies, exceeding 100\% around $\log M_*/M_{\odot} \sim 11.5$. In comparison, upon computing the large-scale (halo-wide) disruptive feedback efficiency, shown in Fig.~\ref{fig:energetics_halo}, we observe that $\overline{E_{\rm{Jet}}}$ for compact and complex radio AGN reaches at most $2-30\%$ of $U_{\rm bin, halo}$ at the lowest mass scales, and around $0.02-0.2\%$ at the highest mass scales.

\subsection{Preventative kinetic feedback in halos}
\label{results:prev_feedback}

We also compare $\overline{E_{\rm{Jet}}}$ to the thermal (cooling) energy, $E_{\rm th}$, of the hot gas in the host halo, to measure the extent to which the cooling processes within the gaseous halo can be offset by jet heating, thus preventing the accumulation of fresh cold fuel for star-formation. $E_{\rm th}$ (as a function of halo mass) is obtained by multiplying the bolometric X-ray [$0.01-100$~keV] luminosity, $L_{\rm bol}$ ---the total X-ray emission of hot gas, predominantly from thermal bremsstrahlung, recombination and two-photon decay--- by the look-back time between $0.0<z<0.4$. $L_{\rm bol}$ as a function of halo mass is derived from the scaling relations\footnote{Note that the $k_bT-M_h$ relation uses $M_{500c}$ (total mass within $r_{500c}$), so we again convert $M_{200c}$ to the required $M_{500c}$. We also scale the resulting $L_{\rm bol}$ to the correct cosmology and redshift interval via $E(z)=H(z)/H(z=0)$, where $H$ is the Hubble parameter.} presented in Figs.~3 and 4 of \citet{Lovisari2021}: $L_{\rm bol} \propto T^{3.04}$ and $M_{\rm 500c} \propto T^{1.76}$, where $T$ is the intracluster medium (ICM) temperature. 

If the total integrated kinetic energy injection is in excess of the halo thermal energy, it is conceivable that the gas could be maintained in a hot state, preventing current (and future) star formation. Therefore, we define this ratio as `preventative kinetic feedback efficiency', $\mathcal{F}_{\rm P}=\overline{E_{\rm{Jet}}}/E_{\rm th}$. Note that the scaling relations derived by \citet{Lovisari2021} are fit to observed groups and clusters with $M_h>10^{13}~M_{\odot}$, so their application to the galaxy regime is merely an extrapolation, and one should be cautious not to over-interpret the results in this regime.

\begin{figure}[t!]
\centering
\includegraphics[width=\linewidth]{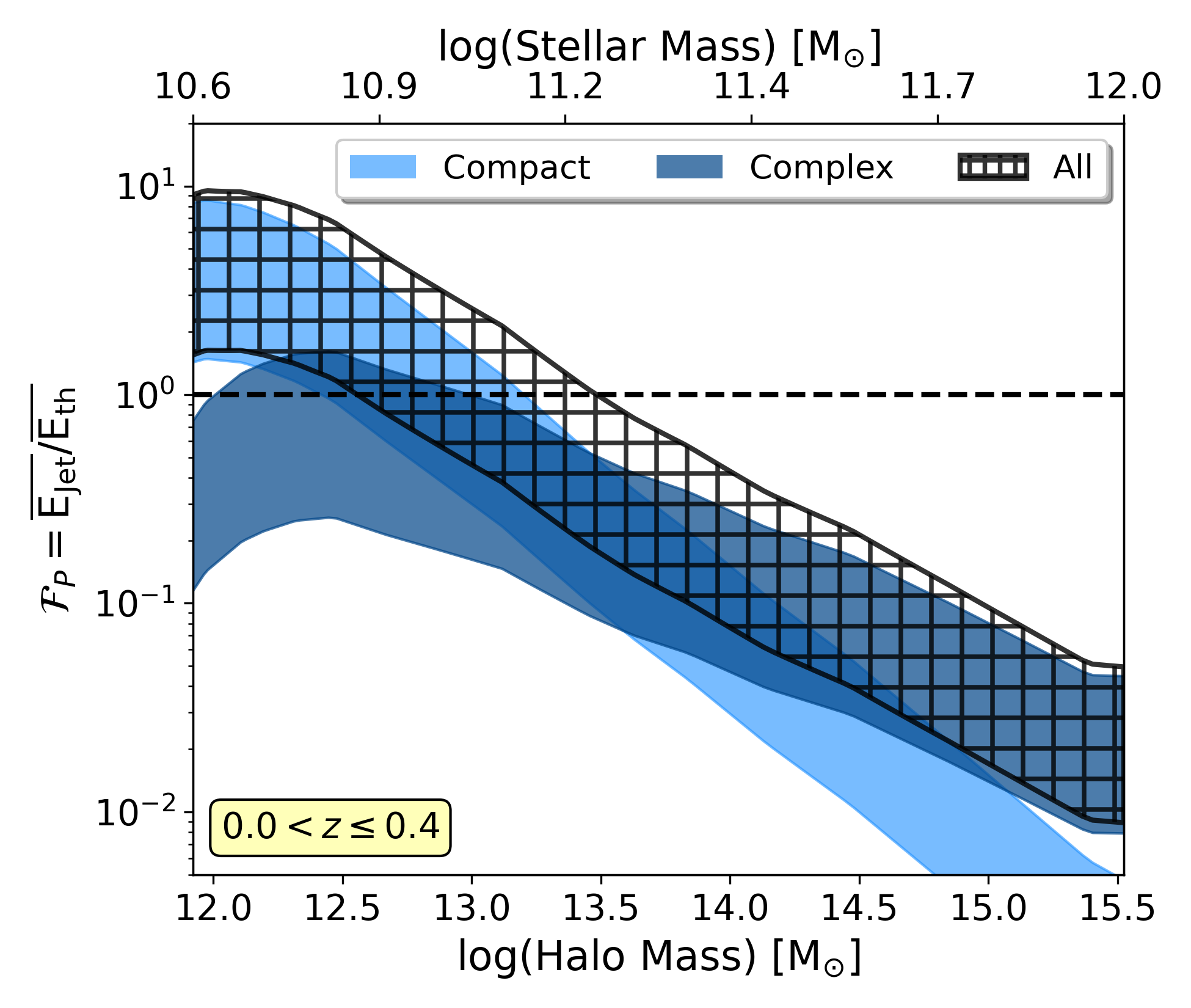}
\caption{Preventative kinetic feedback efficiency, i.e. the ratio of the average jet kinetic energy to the thermal (cooling) energy, as a function of halo mass for the massive galaxy population (black hatched), with respective contributions from the compact and complex radio AGN in light and dark blue shaded regions. The horizontal black dashed line marks the equality between the heating provided by the AGN and the cooling of the halo gas.}
\label{fig:prev_feedback}
\end{figure}

Figure \ref{fig:prev_feedback} shows $\mathcal{F}_{\rm P}$ as a function of halo (and stellar) mass for the massive galaxy population (black hatched), with respective contributions from compact (light blue) and complex (dark blue) radio AGN. The horizontal black dashed line marks the equality of the jetted kinetic energy and the thermal cooling energy. We deduce that, for the galaxy and small group regime, the jet kinetic energy exceeds the total thermal energy of the cooling gas in the halo, and so the preventative feedback effect of the collective population of radio AGN affects the global thermodynamics of these halos. On the other hand, for the large group and cluster regime ($\log(M_h/M_{\odot})>13.5$), $\mathcal{F}_{P}$ drops to $\sim 3-60\%$, suggesting that radio AGN do not inject sufficient energy to globally affect those larger halos.

\begin{figure}[ht!]
\centering
\includegraphics[width=\linewidth]{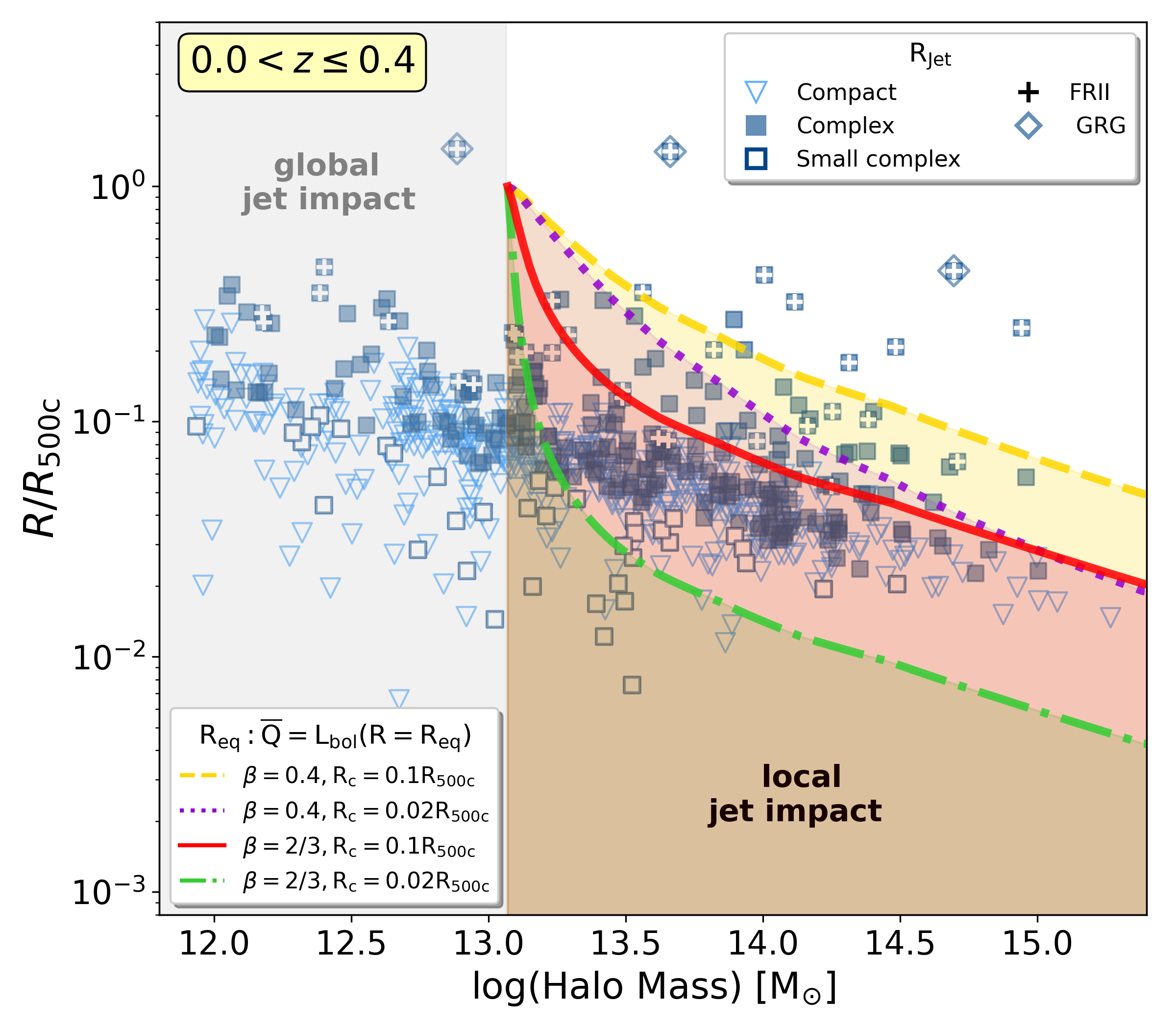}
\caption{Curves showing the `equivalence radius' ($R_{\rm eq}$), the radius at which the integrated thermal cooling luminosity equals the average jet power, as a function of halo mass, normalised by $R_{\rm 500c}$. Red, yellow, green and purple curves denote $R_{\rm eq}$ for a set of $\beta$-profiles with canonical parameters: $\beta=0.4, 2/3$, $R_c=0.02R_{\rm 500c},0.1R_{\rm 500c}$. Given that the definition of $R_{\rm eq}$ requires $R_{\rm Jet} \leq R_{\rm eq}$, the region below each curve (shaded) denotes the parameter space where the radio jets can exert a significant impact on the \textit{local} thermodynamical heating and cooling balance of the halo gas. The different subsets of G9 radio AGN from \citet{Igo2024} are also shown with markers defined in the legend (see text for details).}
\label{fig:jet_impact}
\end{figure}

Finally, in order to explore the local impact of preventative kinetic feedback, we compute the `equivalence radius', $R_{\rm eq}$, defined as the radius at which the integrated thermal cooling luminosity equals that of the average jet power: $\overline{Q}=L_{\rm bol}(R=R_{\rm eq})$. To do so, we express the X-ray surface brightness (SB) profile of the circumgalactic medium of halos as a function of radius ($R$) using a $\beta$-profile \citep{Cavaliere1976}:
\begin{equation}
\label{eq:Sbeta}
     S_{\rm X}   = S_{\rm X,0} \left[  1 + \left(\frac{R}{R_{\rm c}} \right)^2 \right]^{-3\beta + \frac{1}{2}},
\end{equation}
where $S_{\rm X,0}$ is the central SB normalisation (set by $\int_0^{R_{\rm 500c}}~ S_{\rm X}~ 2 \pi R~ dR = L_{\rm bol}$), $R_c$ is the core radius, and $\beta$ is the power-law slopes outside $R_c$. An important assumption in this computation is that the average jet kinetic energy is fully contained within the spherical volume of radius $R_{\rm eq}$. Additionally, the value of $R_{\rm eq}$ depends on the assumed surface brightness profile. Given that there is currently no clear consensus in the literature about a `universal SB profile' especially in the inner regions where non-gravitational processes like AGN feedback can be present and are time-dependent \citep[see e.g ][Eckert et al. in prep.]{Vikhlinen2006, Arnaud2010, Hudson2010, McDonald2014, Ghirardini2019, Kaefer2019, Lehle2024}, the simple but flexible $\beta$-model is deemed adequate to make global energetics statements in the context of this work. Figure \ref{fig:jet_impact} presents several $R_{\rm eq}/R_{\rm 500c}$ curves, derived using varying canonical SB profiles for groups and clusters, with the following parameter values: $\beta=0.4, 2/3$ and $R_c=0.02R_{\rm 500c},0.1R_{\rm 500c}$. These choices are discussed further in Section \ref{sec:discussion}.

We also compute the `normalised jet impact radius' of the sample of `G9 radio AGN' from \citet{Igo2024}, by dividing the (projected) physical jet radii ($R_{\rm Jet}$), defined\footnote{\texttt{LOFAR\_Maj} is the FWHM of the major axis of the source, in degrees. For sources with distant multi-components, the physical size is determined as the largest linear size, as described in \citet{Igo2024}.} as \texttt{LOFAR\_Maj}/2 [in units of kpc], by $R_{\rm 500c}$. Figure \ref{fig:jet_impact} shows this `normalised jet impact radius' as a function of halo mass, for the compact (light blue upper limits), complex (dark blue filled squares) and `small' complex (white squares with dark blue edges) radio AGN, as well as for the subset of FRIIs (white crosses) and giant radio galaxies (GRGs; blue diamonds). $R_{\rm Jet}$ for compact radio AGN is an upper limit on the true physical size of the jets as the source is unresolved at the LOFAR-eFEDS 8\arcsec\ $\times$ 9\arcsec\ resolution. Similarly, we define a complex source to be `small' if it has a physical size $<40$~kpc \citep[i.e. smaller than typical massive galaxy scales][]{vanderWel2014}, roughly corresponding to the resolution limit. Sources which have $R_{\rm Jet}/R_{\rm 500c}\sim 1-5\%$ are at very low ($z<0.1$) redshifts.

Overall, taking the canonical SB profile with $\beta=2/3$ and $R_{c}=0.1R_{\rm 500c}$ for the purpose of this discussion (used for computing the red curve on Fig.~\ref{fig:jet_impact}), $R_{\rm eq}$ reaches of order 10\% and 1\% of the halo $R_{\rm 500c}$ for groups and clusters, respectively. This confirms that, on average, the jet heating in these more massive systems cannot offset the total thermal cooling energy, as the `equivalence radii' of the jets are simply too small compared to the full size of the dark matter halo. On the other hand, the parameter space where $R_{\rm Jet} \leq R_{\rm eq}$ (red shaded region), i.e. where the jetted energy is fully contained within the sphere with radius $R_{\rm eq}$, defines a region where the jets can still exert significant \textit{local} impact on the thermodynamical heating and cooling balance. In fact, the majority of the `G9 radio AGN' in this halo mass regime lie in the region marked by $R_{\rm Jet} \leq R_{\rm eq}$, meaning that jet heating is able to offset cooling flows in the cores of large groups and even the most massive clusters (see \S \ref{disc:glob_energetics} for more discussion).

\section{Discussion}
\label{sec:discussion}

\subsection{On the interpretation and determination of $\overline{Q}(M_*)$}
\label{discussion:caveatsQ}

It is commonly assumed that it is the large, complex radio sources that exert the most powerful feedback on their surroundings. However, when considering the global energy budget from all radio AGN (Fig.~\ref{fig:energetics_Qav}), it is the numerous compact radio AGN, which are often, but not exclusively, less luminous in the radio (recall incidence, Fig.~\ref{fig:compact_complex_incidence}, and RLF, Fig.~\ref{fig:RLF}), that dominate the average jet power for all but the most massive galaxies ($\log M_*/M_{\odot} < 11.5$). 

Generally, $\overline{Q}$ scales with $M_*$ roughly as a power law with index $\sim 2.5$ (or $\overline{Q} \propto M_h$, for our choice of SHMR). Such a relation is similar in slope to that found in the semi-analytical models (SAMs) of \citet{Somerville2008}, but is around 1-2 orders of magnitude lower in normalisation (given our simplistic conversion of $M_{\rm BH}=0.002M_*$). However, our normalisation agrees well with past observational work by \citet{Allen2006} and \citet{Best2006} \citep[see Fig.~11 in][for a summary]{Somerville2008}. Importantly, \citet{Allen2006} conduct detailed \textit{Chandra} X-ray spectral analysis to determine the `Bondi' accretion rates (using also galaxy velocity dispersion to estimate black hole masses) and jet powers from cavity measurements of nine nearby, X-ray luminous elliptical galaxies, and find that a tight correlation exists between these two parameters. As a result, the authors conclude that the `Bondi' formalism, the one usually implemented in SAMs, provides a suitable description of the accretion mechanisms present in their sample of luminous elliptical galaxies. Thus, our empirical $\overline{Q}(M_*)$ measurement may help constrain the radio mode feedback efficiency in SAMs and simulations \citep[e.g. $\kappa_{\rm radio}$, $\kappa_{\rm R}$ in][respectively]{Somerville2008, Croton2016}, a free parameter which is usually fixed to match typical AGN-galaxy observational trends.

As mentioned in Section \ref{sec:LFs}, the most important caveat of this work remains the uncertainty in the determination of jet power from the observable radio luminosity via Eq.~\ref{Qeq}. We refer the reader to Section 6.2.1 of \citet{Igo2024}, where we discuss this issue in detail. However, we note here that the difference in the normalisation of $Q-L_R$ relation \citep[typically by the `uncertainty factors' denoted by $f_W$ or $f_{\rm cav}$, detailed for example in][]{HeckmanandBest2014} can explain the discrepancy between the $\log \Omega_{\rm kin}$ derived by \citet{Hardcastle2019} and the results of this work and of \citealt[][]{Smolcic2017} (see their Figure 6 showing the effect of $f_W$ on $\log \Omega_{\rm kin}$).

The second most important caveat is the unknown origin of radio emission in low-luminosity compact radio AGN \citep[see the recent review of][and references therein]{Panessa2019}. In this work, we assume that the radio luminosity observed in our sample of radio AGN, after thorough cleaning from possible star-formation related emission \citep[see Fig. 9 from][]{Igo2024}, is dominated by (unresolved) jetted emission. However, for those unresolved compact sources, the radio emission may originate from shocks, wind or accretion coronae \citep[which in turn may also be viable feedback mechanisms; e.g.][]{Brinkmann2000, Laor2008, Zakamska2014,Panessa2019, Kawamuro2022}. Eq.~\ref{Qeq} is most probably not valid for these emission processes, and their scaling with black hole mass, stellar mass, or other physical properties, is likely different from jetted emission. Nevertheless, this is only important for the lowest radio luminosity sources as past a certain threshold only a jetted origin is energetic enough to explain the radio emission. 

A further minor uncertainty is the possible redshift evolution of the SMF and incidence within the finite redshift bins, resulting in an evolution of the duty cycle. However, there is only minor redshift evolution from $0<z<0.285$ and $0.285<z<0.4$ (see Figs.~\ref{fig:compact_incidence_fit_results} and \ref{fig:complex_incidence_fit_results}) and this is taken into account by the weighted average to obtain $\overline{Q}$; furthermore, the SMF does not significantly evolve over $0<z<0.4$ \citep[see Fig. 5 from][]{Ilbert2013}.

Lastly, since low frequency LOFAR observations are able to detect radio emission from older electron populations, we cannot easily distinguish active versus remnant jets (without the use of radio data at other frequencies), especially in compact sources. This, therefore, may hamper the interpretations of the measured kinetic energy as a tracer of current feedback in galaxies (e.g. recently triggered or quenched star-formation), as it is likely a result of the \textit{cumulative} jetted events across the lifetime of the source (potentially even from numerous triggering episodes).

\subsection{Importance of the radio luminosity function synthesis}
\label{discussion:rlf}

Observed luminosity functions are often used as metrics to calibrate hydrodynamic and semi-analytic simulation outputs. Yet, such simulations can have vastly different AGN accretion and feedback (`sub-grid') prescriptions and still successfully reproduce the same LFs due to degeneracies in the different fine-tuned input parameters \citep[e.g.][]{Vogelsberger2014, Schaye2015, Croton2016, Kaviraj2017, Springel2018, Dave2019, Habouzit2021}. In addition, it is particularly difficult to predict radio continuum emission due to the complex physics and computing power limitations for resolving the sub-pc to pc scales involved (e.g. evolving the distribution of electrons and the magnetic fields self-consistently), and simultaneously simulating large volumes. Often such radio predictions are added in post-processing using empirical relations between the (known) black hole accretion rates and the (unknown) radio luminosities, for example in recent work by \citet{Slyz2015, Thomas2021} who manage to reproduce the observable RLF from Horizon-AGN and SIMBA cosmological hydrodynamical simulations, respectively. Nevertheless, these works highlight the still uncertain parameter choices in their prescriptions and the lack of physically motivated radio AGN model to couple jets with different accretion modes self-consistently \citep[][see also \citealt{Raouf2017}]{Thomas2021}.

Although our work cannot provide a detailed physical understanding of the mechanisms of jet powering and how exactly this ties into all accretion mode, galaxy and environmental properties, the mass-dependent jet powering mechanism in massive galaxies, as seen from the incidence distributions, is a key ingredient for understanding kinetic AGN feedback. Moreover, decomposing the RLF into stellar mass and radio morphology classes may be helpful to disentangle degeneracies in AGN feedback simulations. 

Figure \ref{fig:RLF} also highlights the need for more deep and large-area radio surveys to probe the low and high luminosity end of the RLF, respectively. This would help in constraining the turnover of the radio AGN incidence at $ \log \lambda_{\rm Jet}<-4$, a key physical constraint to understand the triggering of jets in the radiatively inefficient accretion mode. 

Future work on larger samples, that allow for a separation of the RLF and global energetics into quiescent and star-forming host galaxy subsets, will be essential in understanding the effect of star formation on jet powering \citep[see also discussion about HERGs/LERGs in][and \citealt{Thomas2021} for HERG/LERG RLF from SIMBA simulations]{BestandHeckman2012}.  \citet{Igo2024} already investigated the incidence of radio AGN in different host galaxy types, finding that the fraction of quiescent galaxies hosting radio AGN was similar to that of star-forming galaxies. However, lack of statistics prevented more detailed conclusions to be drawn, but given the differences in the SMF of quiescent and star-forming galaxies \citep{Moustakas2013}, deeper multiwavelength data on larger survey fields might reveal interesting conclusions.

\subsection{On the interpretation of the global energetics}
\label{disc:glob_energetics}

As we have shown above, kinetic feedback dominates over radiative feedback in the local universe for massive galaxies. This is a direct consequence of the total kinetic energy input of both compact and complex radio AGN sources scaling more steeply than the average radiative energy of AGN with galaxy mass (which follows $\overline{E_{\rm rad}}\propto M_*$, due to the near mass-independent X-ray AGN incidence \citealt{Aird2012}), meaning that the `radiative' mode of feedback only begins to dominate at low stellar masses, $\log M_*/M_{\odot} < 10.6$ (see purple dot-dashed line in Fig.~\ref{fig:energetics_Qav}). In fact, this agrees well with recent work of \citet{Petter2024}, who find that powerful jet heating significantly dominates over quasar winds for $M_h\gtrsim10^{13}~h^{-1}~M_{\odot}$, i.e. in the group and cluster regime (at $z<2$, for $h= H_0/100$~km s$^{-1}$Mpc$^{-1} = 0.6766$, see their Fig. 11). This is interesting as they use a completely independent method to derive the clustering and halo masses via halo occupation distribution (HOD) modelling for sample of radio AGN from \citet{Best2023}, and later use this information to derive the energetics. 

Our result also agrees with recent observational work by \citet{Buchner2024, Kondapally2023, HeckmanBest2023}. \citet{Buchner2024} combines distribution functions and scaling relations to derive average outflow rates as functions of mass and cosmic time, concluding that massive galaxies at $z<0.3$ are predominately prevented from growing further by jet heating. Similarly, \citet{Kondapally2023} estimate a kinetic heating rate as a function of radio luminosity for different subsets of the radio AGN population (including quiescent low-excitation radio galaxies and all radio-excess AGN), as well as comparing to a suite of semi-analytical and hydrodynamical simulations, finding that AGN jets play a dominant role in AGN
feedback at $z \lesssim 2$. These results are further supported by the work of \citet{HeckmanBest2023}, who compare the energy injection from massive stars and supernovae, radiation pressure and winds driven by AGN, and AGN radio jets, finding that the amount of $E_{\rm kin}$ for jets is an order of magnitude larger than from AGN winds at least up to $z \sim 1$. The authors further show that the maximum kinetic energy injection by jets occurs around $z \sim 1$, a lower redshift than the peak of star-formation and radiative AGN activity (`cosmic noon', $z \sim 2-3$). They derive a time-integrated (i.e. across the entirety of cosmic history) total kinetic energy per unit volume due to jets of $U_{\rm Jet}= 2.6 \times 10^{57}$~erg~Mpc$^{-3}$. Taking simply our local universe estimate of $\log \Omega_{\rm kin}/[\rm {W~Mpc^{-3}}] \sim 32.15$, propagated across the entirety of cosmic history, we derive a value of $U_{\rm Jet}=6.1 \times 10^{56}$~erg~Mpc$^{-3}$, which is of similar order (albeit lower as the increase of $\Omega_{\rm kin}$ up to $z\sim1$ is not considered in the calculation).

In terms of the small-scale `disruptive' feedback, we showed in Fig.~\ref{fig:energetics_gal} that the compact radio AGN do not have enough jet kinetic energy to surpass the binding energy of galaxies, but $\overline{E_{\rm Jet}}$ reaches significant fractions of $U_{\rm bin, gal}$ for higher and higher stellar masses \citep[see also][]{HeckmanBest2023}. This means that, although energetically the gas cannot be unbound from such galaxies, it may be significantly disrupted in its kinematics and distribution, potentially impacting both the star formation and the central gas supply for fuelling the AGN \citep[e.g.][]{McNamara2014,Morganti2015}. Factors that can affect the extent of gas disruption include the morphology, collimation, entrainment and mass-loading of the jet \citep[][O'Shea et al. in prep.]{DeYoung1986, Bicknell1986, Bowman1996, Hubbard2006}, as well as the structure of the surrounding ISM or inter-group medium \citep[IGrM; e.g.][]{Tanner2022, Dutta2024,HardcastleKrause2013, English2016, Croston2019, Gaspari2020, Morris2022,  Mingo2022}, but further detailed discussion is out of the scope of this work. As for the complex sources, $\mathcal{F}_{\rm D, gal}>1$ for the highest masses, $\log (M_*/M_{\odot}) > 11.5$. However, from the analysis in Section \ref{results:prev_feedback}, around 86\% of complex LOFAR radio AGN have physical sizes (at the LOFAR-eFEDS resolution) greater than $\sim 40$~kpc \citep[see also Fig.~23 in][]{Igo2024}, extending beyond typical galaxy scales \citep{vanderWel2014}. This suggests that a large fraction of the jet kinetic energy (depending on some of the factors mentioned above) may be deposited outside the galaxy, and therefore a comparison with the galactic binding energy may not provide physically meaningful results for all the complex radio AGN.

This is why we also compute the large-scale disruptive feedback efficiency (Fig.~\ref{fig:energetics_halo}), which shows that the average jet kinetic energy of both compact and complex radio AGN is largely insufficient at all mass scales to unbind all the gas from the host halo: the binding energy of dark matter halos is simply too large compared to the kinetic energy that radio AGN (at low redshift) can provide. At galaxy and group scales $\mathcal{F}_{\rm D,~halo}$ can reach up to $\sim 30\%$ (considering a mass-dependent $f_{\rm gas}$), but a steep decline with $M_h$ means that for the cluster regime $\mathcal{F}_{\rm D,~halo}$ is only $\sim 0.2\%$. This is in line with the universal gas fractions observed in clusters \citep[whereas groups are preferentially gas-depleted; e.g.][]{Eckert2021}, as the deep potential wells maintain the primordial ratio of cosmological baryon-to-total-matter densities.

Thus far, we have shown that, \textit{globally}, radio AGN in the local universe do not exert enough small- and large-scale disruptive feedback to their host galaxies and halos. Nevertheless, radio AGN may be an important source of `preventative' feedback as shown by Figures \ref{fig:prev_feedback} and \ref{fig:jet_impact}. Given that $U_{\rm bin} \propto M_{h}^2/R_{\rm 200c} \sim M_{h}^{1.67}$ and $L_{\rm bol} \propto M_{h}^{1.73}$ \citep{Lovisari2021}, $U_{\rm bin}$ scales quasi-linearly with $L_{\rm bol}$, i.e. with the total thermal energy of hot halos.
Therefore, $\mathcal{F}_{\rm P}$ follows a similarly steep evolution with $M_h$ as $\mathcal{F}_{\rm D,~halo}$, but now the normalisation is higher, as shown in Fig.~\ref{fig:prev_feedback}, reflecting the simple fact that for massive halos, the thermal energy of the hot gas is smaller than their binding energy. 

In the galaxy and small group regime, we measure $\mathcal{F}_{\rm P} \gtrsim 1$, meaning the jet energy can be greater than the $E_{\rm th}$. This suggests that the heating provided by the jet can efficiently offset the cooling in these smaller halos. In the large group and cluster regime, on the other hand, $\overline{E_{\rm jet}}$ reaches only $\sim 3-60\%$ of the total thermal energy of the cooling gas in the halo, suggesting that radio AGN cannot impact the global thermodynamical equilibrium of these systems. This can be explained by Fig.~\ref{fig:jet_impact}, where we show that $R_{\rm eq}/R_{\rm 500c}$ declines from of order 10\% at group scales to of order 1\% for cluster scales, meaning that as the halo mass increases, the radial impact of the jets decrease \citep[see also e.g.][]{Eckert2021}. 

However, even though the reach of the jet is small compared to the virial radius of the halo, we know from X-ray observations of clusters \citep[and simulations, e.g][]{Croton2006, Bower2006,Somerville2008,Croton2016} that preventative feedback is a necessity to prevent `catastrophic cooling flows' \citep{Fabian1994, Peterson2004, McNamara2012}, which would be expected given the short radiative cooling timescales compared to the cluster ages. In fact, taking again the canonical SB profile with $\beta=2/3$ and $R_{c}=0.1R_{\rm 500c}$ as an example, most G9 radio AGN lie in the region where $R_{\rm Jet} \leq R_{\rm eq}$. Therefore, low-redshift radio AGN living in groups and even in the most massive clusters, do have the power to exert significant \textit{local} impact on the thermodynamical balance in the central cores of their host halos. Interestingly, it is mainly the FRII and GRG sources that preferentially lie above the red curve ($R_{\rm Jet} > R_{\rm eq}$), which is not surprising given that the spatial distribution of their jet kinetic energy is edge- rather than core-brightened, meaning that they deposit the bulk of their energy further from the halo core. This, therefore, could impact their ability to effectively offset strong central cooling flows, and thereby also affect the fuelling of the central AGN, although this remains to be further investigated, as the feeding and feedback cycle is complex and time-dependent \citep[e.g.][]{Gaspari2020, Mingo2022}.

Moreover, the $R_{\rm Jet}$ distribution of the observed radio AGN traces very well the trends of the red canonical $R_{\rm eq}$ curve, clustering just below the curve. This may be a coincidence of the given LOFAR resolution, or be a potential indication of the natural heating and cooling equilibrium reached by the jet and the inner regions of the gaseous halo. Although a comparison of individual realisations of jetted AGN (the data points), which may be variable in time, and population-averaged quantities (the curves) should be done with caution. Future high resolution LOFAR VLBI studies \citep{Morabito2022, Sweijen2022}, capable of resolving the detailed jet structures on large samples of radio AGN, will be needed to shine light on this matter. 

Past works have also made local and global statements about the heating-cooling balance in the environments of radio AGN. For example, \citet{Hardcastle2019} integrate the Schechter profile of the local cluster luminosity function obtained by \citet{Boehringer2014} to get a cooling luminosity of $2 \times 10^{31}$~W~Mpc$^{-3}$ and conclude that the derived heating rate (see green star on Figure \ref{fig:kond2023}) from their sample of 23,344 LOFAR radio AGN can offset (in statistical terms) the radiative cooling in these systems \citep[see also][]{Smolcic2017, Butler2019, Croton2016}. Similarly, \citet{Dunn2006} use detailed X-ray and radio observations of X-ray cavities and spatially coincident radio bubbles to conclude that the average radius to which radio bubbles could offset X-ray cooling ($r_{\rm heat}$) was $r_{\rm heat}/r_{\rm cool}=0.86 \pm 0.11$, (where $r_{\rm cool}$ is the radius within which the cooling time equals 3~Gyr) and 10/16 clusters had $r_{\rm heat}/r_{\rm cool} \gtrsim 1$ \citep[see also][for a review]{McNamaraNulsen2007}.

\subsection{On jet energy deposition efficiency, cluster profiles and halo occupation distributions}

Throughout this work, we assume that 100\% of the jet kinetic energy couples to the surrounding medium, via thermal dissipation, sound waves, shocks, turbulence, release of cosmic rays and other mechanisms which can also indirectly transfer the jet energy to the ISM/CGM \citep[e.g][and references therein]{DeYoung1986, Bicknell1986, Wagner2012, Wykes2013, Zhuravleva2014, Jacob2017, Perucho2019, Hlavacek-Larrondo2022}. Without going into details of these physical processes, this must be a reasonable assumption if the jet fully decelerates within the host galaxy/halo. In fact, other than two giant radio galaxies \citep[see Appendix C of][]{Igo2024}, all G9 radio AGN, from which the energetics results in this work are computed, have $R_{\rm Jet}/R_{\rm 500c}<1$, so the aforementioned assumption is reasonable, at least when considering the (large) halo scales. We also note that the distinction between preventative and disruptive feedback is somewhat artificial and the total jet kinetic energy is not distributed in a mutually exclusive way between the two; the physics of these two processes are closely interlinked.

Regarding Figure \ref{fig:jet_impact}, 
our choices for $\beta=0.4,2/3$ are justified as they represent well the ranges of SB profile slopes found in past literature \citep[e.g. Sanders et al., A\&A (2025) accepted, ][and references therein]{Vladutescu-Zopp2024}. The range of core radii is also as of yet uncertain, and can be degenerate with $\beta$. We chose to present $R_c$ in the range of $0.02-0.1R_{\rm 500c}$ as these are some canonical values for cool-core and non-cool core clusters \citep[e.g.][]{Wang2023}. However, it seems that the combination of $\beta=2/3$ and $R_{c}=0.02R_{\rm 500c}$ (green curve) represents a SB profile that is unphysically centrally concentrated, as most observed radio AGN jets lie above the green curve. Using a more complex \citet{Vikhlinen2006} model for the SB, with variable inner ($R<R_c$) power-law slope, $\alpha$, would also show similarly low radii of equivalence for $\alpha>0$, as more of the luminosity would be concentrated in a smaller volume.

Interestingly, \citet{Wang2023} quantifies the cool-core condensation radius ($R_{\rm ccc}$: radius within which the cooling time equals the turbulence eddy turnover time\footnote{$R_{\rm ccc}$ is a measure of the balance between feeding and feedback processes, generating turbulent condensation rain and related chaotic cold accretion \citep{Wang2023, Gaspari2018}}) and quenching cooling flow radius ($R_{\rm qcf}$: radius within which the cooling time is 25 times the free fall time\footnote{$R_{\rm qcf}$ encompasses the region of thermally unstable cooling \citep{Wang2023, Voit2015}}) for a range of massive, nearby clusters ($1.3 \times 10^{14}<M_{500c}/M_{\odot}<16.6 \times 10^{14} ; 0.03<z<0.29$). They find typical values of  $0.01 < R_{\rm ccc}/R_{\rm 500c} < 0.05$ and  $0.02 < R_{\rm qcf}/R_{\rm 500c} < 0.13$, both of which span over the observed parameter space of $R_{\rm Jet}/R_{\rm 500c}$ for those radio AGN in the cluster regime. This highlights in an independent manner the effectiveness of jet heating offsetting cooling in the central regions of massive clusters.

Finally, we have taken all radio AGN so far to be central, and not satellite, galaxies in their host halos. Halo occupation distribution (HOD) models predict the number of satellites to increase as a function of halo mass, and it is currently unknown how the occupation fraction of radio AGN changes between centrals and satellites \citep[e.g.][]{Berlind2003, Zheng2005, Comparat2023}. Radio AGN triggering in satellite galaxies is also likely to work differently to centrals due the local distribution and/or dynamics of the gas (e.g gas stripping at the outskirts of clusters), and different formation histories, potentially impacting the incidence distributions and thereby our statements on the global energetics. For example, \citet{Vos2024} (and related simulation work by \citealt{Rihtarsic2024}) find that the observed LOFAR radio AGN fraction peaks near the cluster core (as is expected for reasons described below) but then \textit{declines} before rising again in the cluster outskirts ($\sim$10 $R_{\rm 500}$), potentially due to lower velocity dispersion at these radii allowing for more mergers to occur and potentially trigger radio AGN \citep[see also e.g.][for more discussion on the importance of mergers for radio AGN triggering]{RamosAlmeida2011,RamosAlmeida2012, Pierce2022}. However, \citet{Best2007} clearly showed that brightest group or cluster (BGG, BCG) galaxies have a higher probability to host radio AGN compared to other galaxies of the same mass \citep[see also e.g.][]{Burns1990, Sun2009, Smolcic2011}, likely due to the more effective condensation of cold clouds from the hot halo or direct hot gas fuelling near the centres of the halo \citep{Hardcastle2007, Gaspari2020}. In fact, upon matching the G9 radio AGN sample to the GAMA Groups (\texttt{G3CGalv10}) and Friends of Friends (FoF; \texttt{G3CFoFGroupv10}) catalogues \citep{Robotham2011groups}, 78\% of sources are associated with the BCG. Therefore, treating our radio AGN as (mostly) centrals appears justified. Future work computing the incidence of radio AGN in centrals and satellites separately will be crucial to rigorously test this effect, and provide a more in depth interpretation of Fig.~\ref{fig:jet_impact}.

\section{Summary}
\label{sec:conclusion}

In this work, we have used the radio AGN incidence as a function of specific black hole kinetic power, stellar mass and radio morphology to quantify the average jet power of massive galaxies in the local universe and interpret this in the context of AGN kinetic feedback energy balance on galaxy and halo scales. 

As in \citet{Igo2024}, we show that the incidence of compact and complex radio AGN is mass-dependent, whereby higher mass galaxies are more likely to host radio AGN across the $\lambda_{\rm Jet}$ range, but they follow different distributions (Fig.~\ref{fig:compact_complex_incidence}). The former follows a steep power law distribution with slope $\sim-1.4$ dominated by jets of lower power, whereas the latter follows a double-power law-like distribution with shallower faint-end slope ($\sim 0.2$), but reaching to high jet powers. 

We then synthesise the total radio AGN luminosity function by convolving the radio AGN incidences with the stellar mass function, allowing us to decompose, for the first time, the RLF as a function of stellar mass and radio morphology (Fig.~\ref{fig:RLF}). We find that in the luminosity range $\log (L_{\rm{150MHz}}/[{\rm W~ Hz}^{-1}]) \sim 23.5-25$, radio AGN with $\log (M_*/M_{\odot}) \sim 11.4$ dominate the RLF in number density. We also find that the compact (complex) radio AGN contribute dominantly in the range below (above) $\log (L_{\rm{150MHz}}/[{\rm W~ Hz}^{-1}]) \sim 24.5$, equivalent to a jet power of $\log (Q/[{\rm W}]) \sim 36.7$. Our RLF and its integrated quantities, such as the average kinetic energy density $\Omega_{\rm kin}$, are in good agreement with past literature. 

Importantly, we find that compact radio AGN dominate the average injected kinetic jet power for all but the most massive galaxies, $\log M_*/M_{\odot} < 11.5$, at $z<0.4$. The total kinetic energy released by complex radio sources grows more steeply with stellar mass than that released by compact sources, and both scale more steeply than the average radiative energy (Fig.~\ref{fig:energetics_Qav}). Taking a `canonical' fraction of radiative output from multi-phase AGN winds, $0.5\%~L_{\rm bol}$, we also see that this `radiative' mode of feedback only begins to dominate at low stellar masses, $\log M_*/M_{\odot} < 10.6$. In an integrated sense, we quantitatively show that kinetic feedback dominates over radiative feedback in the local universe (Fig.~\ref{fig:kond2023}). 

We then define three metrics to gauge the efficacy of kinetic feedback to either disrupt the galaxy-wide and halo-wide gas distribution, potentially driving it out of the system completely (`disruptive feedback'), or to prevent the gas from cooling and forming stars (`preventative feedback'). 

We find that compact radio AGN do not have enough jet kinetic energy to efficiently unbind gas from galaxies across the probed mass scale, although they reach a level that could be sufficient to significantly affect the local gas distribution. On the other hand, complex sources show $\mathcal{F}_{\rm D,~gal}>100\%$ for $\log M_*/M_{\odot} \sim 11.5$ (Fig.~\ref{fig:energetics_gal}). Nevertheless, given that the majority ($\sim$ 86\%) of complex sources have physical sizes larger than about $40$~kpc, they extend on average past the stellar body of typical galaxies, meaning that the jet kinetic energy is mostly deposited on larger scales. Therefore, we consider also the efficacy of the halo-wide disruptive feedback and find that at no mass range are jets energetic enough to unbind gas from entire halos (Fig.~\ref{fig:energetics_halo}). Although, $\mathcal{F}_{\rm D, halo}$ does reach $\sim 2-30\%$ at galaxy and group scales, in line with the possibility of reducing fraction of gas in these systems away from the cosmic value. 

Lastly, we show that the jet kinetic energy of AGN may be sufficient to offset the cooling in halos (Fig.~\ref{fig:prev_feedback}). In fact, in the galaxy and small groups regime, $\overline{E_{\rm Jet}}(M_h) \gtrsim E_{\rm th}$, meaning that the kinetic energy can impact the heating and cooling balance of the gas on global scales, i.e. throughout the halos. This is not the case at cluster scales, however, where $\mathcal{F}_{\rm P}$ is just a few per cent. Nonetheless, by comparing the physical sizes of observed radio jets in our complete sample of radio AGN to the radius within which the integrated surface brightness of the halo is equal to $\overline{Q}$, we find that they preferentially populate the region where $R_{\rm Jet} \leq R_{\rm eq}$ (Fig.~\ref{fig:jet_impact}). Acknowledging the caveats present in the assumption of a simple $\beta$-profile for halos across the mass scale, we conclude that jetted AGN feedback can contribute significantly to the \textit{local} heating of the gas in the central cores of groups and even the most massive clusters, where we do indeed observe strong cool-cores and jet-inflated bubbles or cavities.

Overall, we have shown the power of combining  AGN incidence measures, thanks to the well-characterised, complete samples from \citet{Igo2024}, with some simple assumptions about their host galaxies and halos, allowing for meaningful physical statements to be made about the global kinetic energy budget of the local universe. Future work on expanding the samples both to high sensitivity and larger volumes, as well as drawing parallels between observational work and AGN feedback simulations, will be vital to complement this knowledge in currently unattainable parameter/simulation space.

\begin{acknowledgements}
    The authors thank the anonymous referee for their careful reading of the paper and their constructive comments. 
    ZI acknowledges the support by the Excellence Cluster ORIGINS which is funded by the Deutsche Forschungsgemeinschaft (DFG, German Research Foundation) under Germany´s Excellence Strategy – EXC-2094 – 390783311. We acknowledge Emre Bahar, Johannes Buchner, Dominique Eckert, Fran{\c c}ois Mernier, R{\"u}diger Pakmor, Paola Popesso, Jeremy Sanders, Soumya Shreeram, Sylvain Veilleux and Tiago Costa for useful discussions. This research was supported by the Munich Institute for Astro-, Particle and BioPhysics (MIAPbP) which is funded by the Deutsche Forschungsgemeinschaft (DFG, German Research Foundation) under Germany´s Excellence Strategy – EXC-2094 – 390783311.
    \\
     
\end{acknowledgements}

\bibpunct{(}{)}{;}{a}{}{,} 

\bibliographystyle{aa} 
\bibliography{aa52888-24.bib} 

\appendix

\section{Modelling the mass-dependence of the radio AGN incidence}

\begin{figure}[h!]
\centering
\includegraphics[trim=1cm 1cm 1.8cm 0.5cm,clip,width=\linewidth]{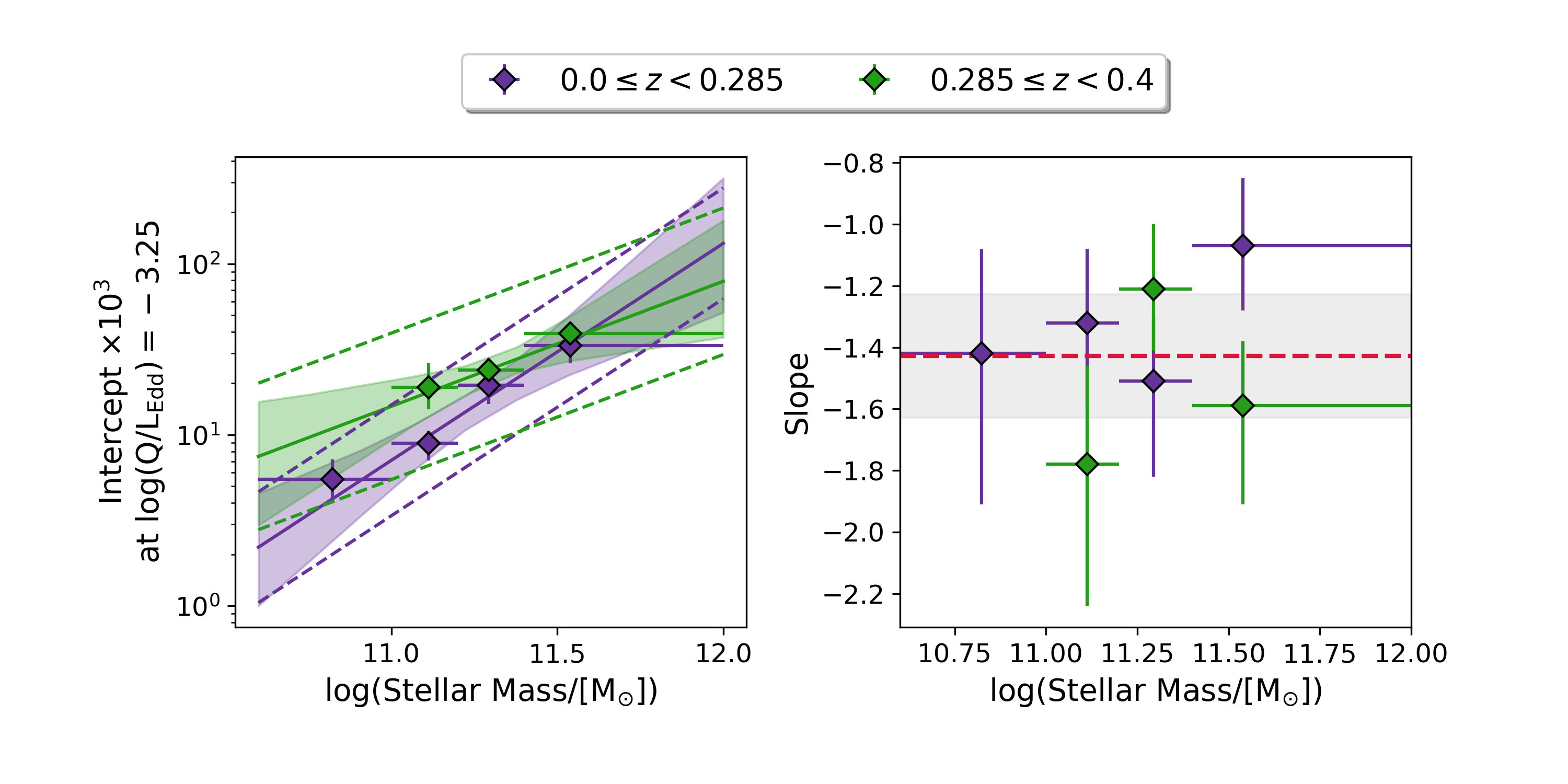}
\caption{Left: Increase of the normalisation of the compact radio AGN incidence found as a function of stellar mass for the low (purple) and high (green) redshift bins. Solid lines with $1\sigma$ shaded error regions mark the best-fit and dashed lines mark the intrinsic scatter. Right: Approximately constant steep slope of $-1.4$ (red dashed line) with standard error of $0.2$ (grey shaded region) found as a function of stellar mass and redshift.}
\label{fig:compact_incidence_fit_results}
\end{figure}

\begin{figure}[h!]
\centering
\includegraphics[width=\linewidth]{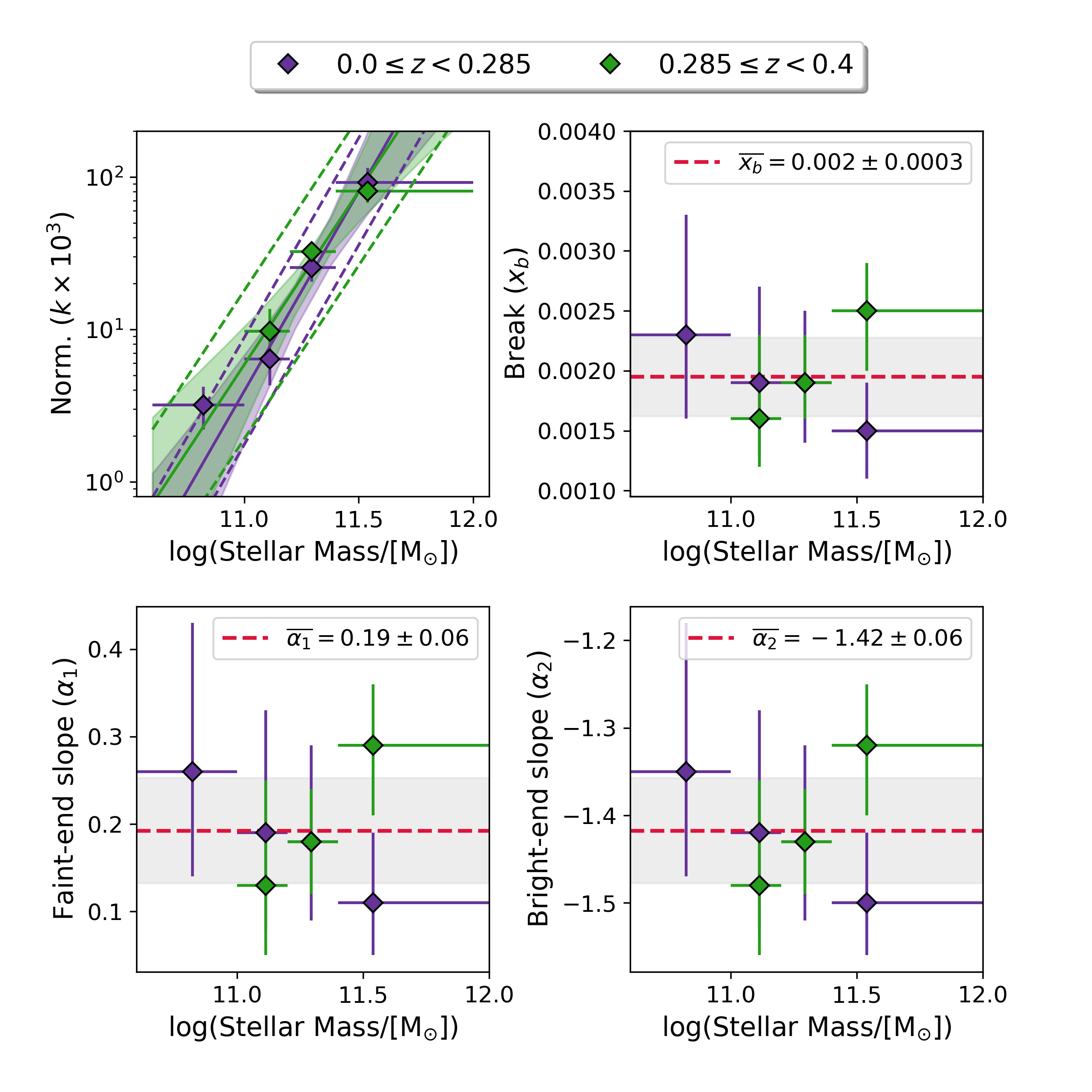}
\caption{Left: Increase of the normalisation of the complex radio AGN incidence found as a function of stellar mass for the low (purple) and high (green) redshift bins. Solid lines with $1\sigma$ shaded regions mark the best-fit and dashed lines mark the intrinsic scatter. Right: Approximately constant break, faint- and bright-end slopes (red dashed lines with grey shaded standard error) are found as a function of stellar mass and redshift.}
\label{fig:complex_incidence_fit_results}
\end{figure}

Figures \ref{fig:compact_incidence_fit_results}
and \ref{fig:complex_incidence_fit_results} show the parameter results from the power-law and double power-law fits to the compact and complex radio AGN incidence, respectively, as a function of stellar mass. The average $M_*$ within each bin is used in the fitting instead of the geometric bin centre to reflect the true distribution. 

\section{Total radio AGN incidence distribution}

\begin{figure}[h!]
\centering
\includegraphics[width=\linewidth]{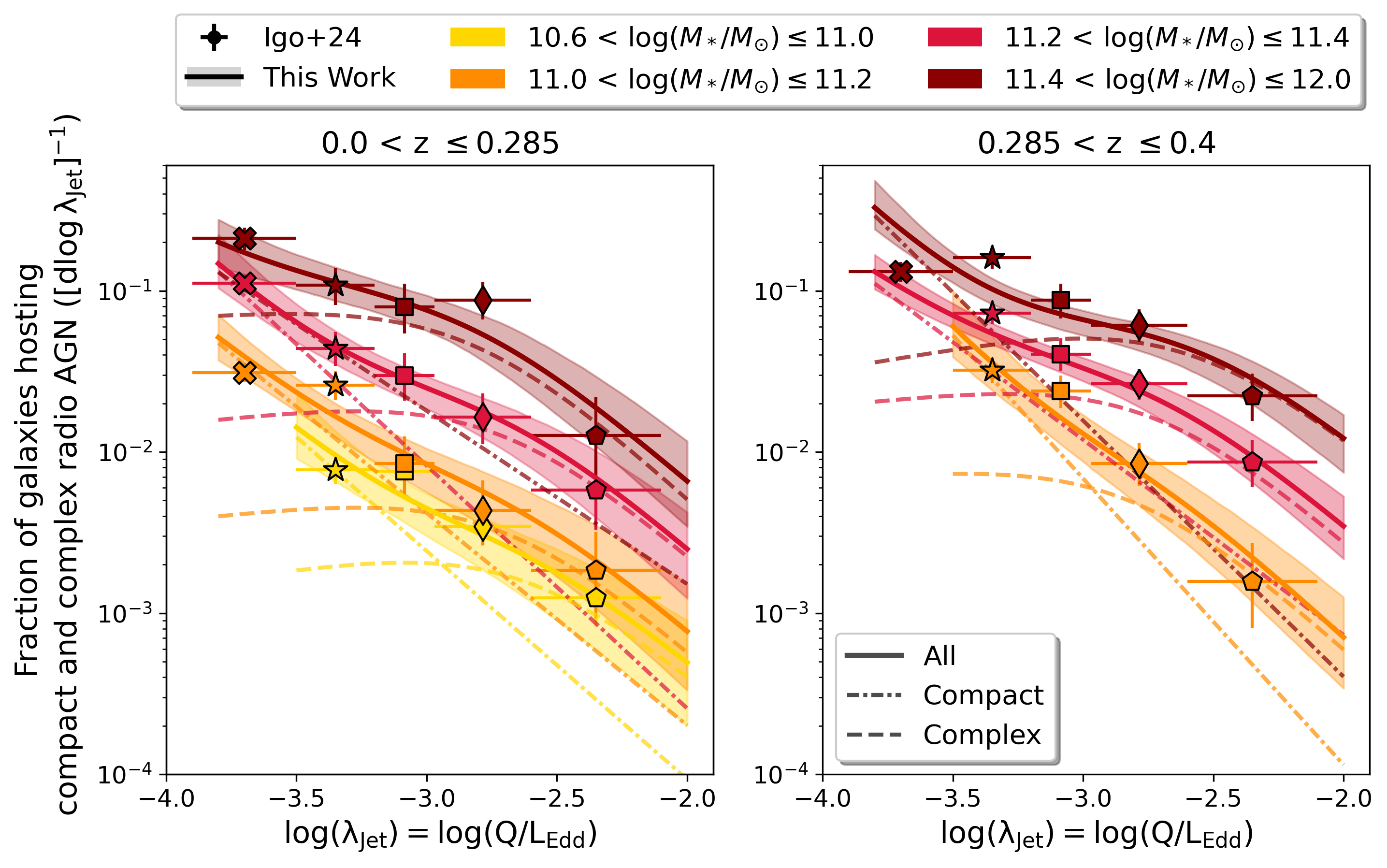}
\caption{Comparison of the incidence of radio AGN as a function of $\lambda_{\rm Jet}$ from \citet{Igo2024} (markers) and this work (best-fit model and error margin), in units of $[\log \lambda_{\rm Jet}]^{-1}$, showing consistency.}
\label{fig:consistent_past_incidence}
\end{figure}

Figure \ref{fig:consistent_past_incidence} shows the sum of the power-law (Eq.~\ref{eq:pow}; Figure \ref{fig:compact_incidence_fit_results}) and double-power-law (Eq.~\ref{eq:doubpow}; Figure \ref{fig:complex_incidence_fit_results}) fits to the total compact and complex incidence of radio AGN (solid curves with shading) as a function of $\lambda_{\rm Jet}$ from this work. The dot-dashed and dashed curves here show the compact and complex contributions, respectively, to the total incidence curves; these are shown in detail in Fig.~\ref{fig:compact_complex_incidence}. Importantly, these results, now represented in units of $[\log \lambda_{\rm{Jet}}]^{-1}$, are fully consistent with \citet{Igo2024} (markers) and are independent of the choice of binning.  

\end{document}